 \documentclass[final,5p,times,twocolumn]{elsarticle}
\usepackage{float}
\usepackage{graphicx}
\usepackage{amssymb}
\usepackage{amsmath}
\usepackage{bm}
\usepackage{physics}
\usepackage{afterpage}
\usepackage{enumerate}
\usepackage{algorithm}
\usepackage{algorithmicx}
\usepackage{algpseudocode}

\usepackage{ulem}

\usepackage[
colorlinks=true,
urlcolor=blue,
citecolor=blue,
linkcolor=blue,
hyperfootnotes=false]{hyperref}

\usepackage{tipa,tipx}

\def\ket#1{ | #1 \rangle }
\def\bra#1{ \langle #1 | }

\newcounter{bla}

\journal{Computer Physics Communications}

\begin{document}

\begin{frontmatter}



\title{TTNOpt: Tree tensor network package for high-rank tensor compression}

\author[a]{Ryo Watanabe\corref{author}}
\author[a]{Hidetaka Manabe}
\author[b]{Toshiya Hikihara}
\author[c,d]{Hiroshi Ueda}

\cortext[author] {Corresponding author.\\\textit{E-mail address:} u293494e@ecs.osaka-u.ac.jp}
\address[a]{Graduate School of Engineering Science, The University of Osaka, 1-3 Machikaneyama, Toyonaka, Osaka 560-8531, Japan}
\address[b]{Graduate School of Science and Technology, Gunma University, Kiryu, Gunma,
Japan}
\address[c]{Center for Quantum Information and Quantum Biology, The University of Osaka, Toyonaka, 560-0043, Japan.}
\address[d]{
Computational Materials Science Research Team, RIKEN Center for Computational Science (R-CCS), Kobe, Hyogo 650-0047, Japan}

\begin{abstract}
We have developed TTNOpt, a software package that utilizes tree tensor networks (TTNs) for quantum spin systems and high-dimensional data analysis.
TTNOpt provides efficient and powerful TTN computations by locally optimizing the network structure, guided by the entanglement pattern of the target tensors.
For quantum spin systems, TTNOpt searches for the ground state of Hamiltonians with bilinear spin interactions and magnetic fields, and computes physical properties of these states, including the variational energy, bipartite entanglement entropy (EE), single-site expectation values, and two-site correlation functions.
Additionally, TTNOpt can target the lowest-energy state within a specified subspace, provided that the Hamiltonian conserves total magnetization.
For high-dimensional data analysis, TTNOpt factorizes complex tensors into TTN states that maximize fidelity to the original tensors by optimizing the tensors and the network.
When a TTN is provided as input, TTNOpt reconstructs the network based on the EE without referencing the fidelity of the original state.
We present three demonstrations of TTNOpt:
(1) Ground-state search for the hierarchical chain model with a system size of $256$.
The entanglement patterns of the ground state manifest themselves in a tree structure, and TTNOpt successfully identifies the tree.
(2) Factorization of a quantic tensor of the $2^{24}$ dimensions representing a three-variable function where each variant has a weak bit-wise correlation.
The optimized TTN shows that its structure isolates the variables from each other.
(3) Reconstruction of the matrix product network representing a $16$-variable normal distribution characterized by a tree-like correlation structure.
TTNOpt can reveal hidden correlation structures of the covariance matrix.

\end{abstract}


\end{frontmatter}

{\bf PROGRAM SUMMARY}
\begin{small}

\noindent
{\em Program Title:} TTNOpt \\
{\em CPC Library link to program files:} (to be added by Technical Editor) \\
{\em Developer's repository link:} Reference \cite{TTNOpt}  \\
{\em Licensing provisions:} Apache 2.0 \\
{\em Programming language:} Python \\
{\em External routines/libraries:} Reference \cite{roberts2019tensornetwork} \\
{\em Nature of problem:} \\
Characterizing the entanglement structure of the lowest energy state of quantum spin systems and of high-dimensional tensor data, for efficient representation. \\
{\em Solution method:} \\
Tensor network contractions combined with a variational algorithm based on the Lanczos method, with automatic structural optimization of tree tensor networks. \\
{\em Restrictions:} \\
Applicable to quantum spin systems and data that can be represented as tensors. \\
{\em Unusual features:} \\
Adaptive structural reconfiguration of TTNs based on the system’s entanglement pattern.
\end{small}

\section{Introduction}
\label{Intro}
Tensors, as multidimensional arrays, are widely used across various computational sciences, including condensed matter physics, big data analytics, and machine learning.
A fundamental difficulty with manipulating tensors is that the number of tensor elements grows exponentially with the tensor rank $N$.
One promising approach to overcoming this challenge is to employ the tensor network (TN) representation (decomposition), in which the tensor of interest is expressed as a contraction of small, low-rank tensors~\cite{orusPracticalIntroductionTensor2014}.
By setting an upper bound $\chi$ on the dimensions of each index (mode), which is referred to as auxiliary bonds, in $O(N)$ small tensors during factorization, the total number of elements in the TN can be reduced to $O(N\chi^p)$ where $p$ reflects the maximum rank of small tensors.
TNs have found broad applications in condensed matter physics and data science. 
In the former, high-rank tensors, such as wave functions and Boltzmann weights, are handled within the TN framework~\cite{okunishiDevelopmentsTensorNetwork2022,orusTensorNetworksComplex2019a}.
In the latter, TNs are utilized for representing complex data, including images~\cite{luTensorNetworksEfficient2024,jobstEfficientMPSRepresentations2024}.
Further expanding their applications are particularly in machine learning~\cite{stoudenmireSupervisedLearningQuantumInspired2017,PhysRevX.8.031012, PhysRevB.99.155131}.
Quantic tensors have been used for enabling the treatment of functions with continuous variables~\cite{oseledetsApproximationMatricesLogarithmic2009,khoromskijOdlogNQuanticsApproximation2011}.
So far, several types of TN structures have been developed, particularly in the quantum many-body physics~\cite{10.1143/PTP.105.409,PhysRevLett.99.220405,PhysRevLett.101.250602, RevModPhys.93.045003}.

\begin{figure}[t]
  \centering
  \includegraphics[clip, width=3.3in]{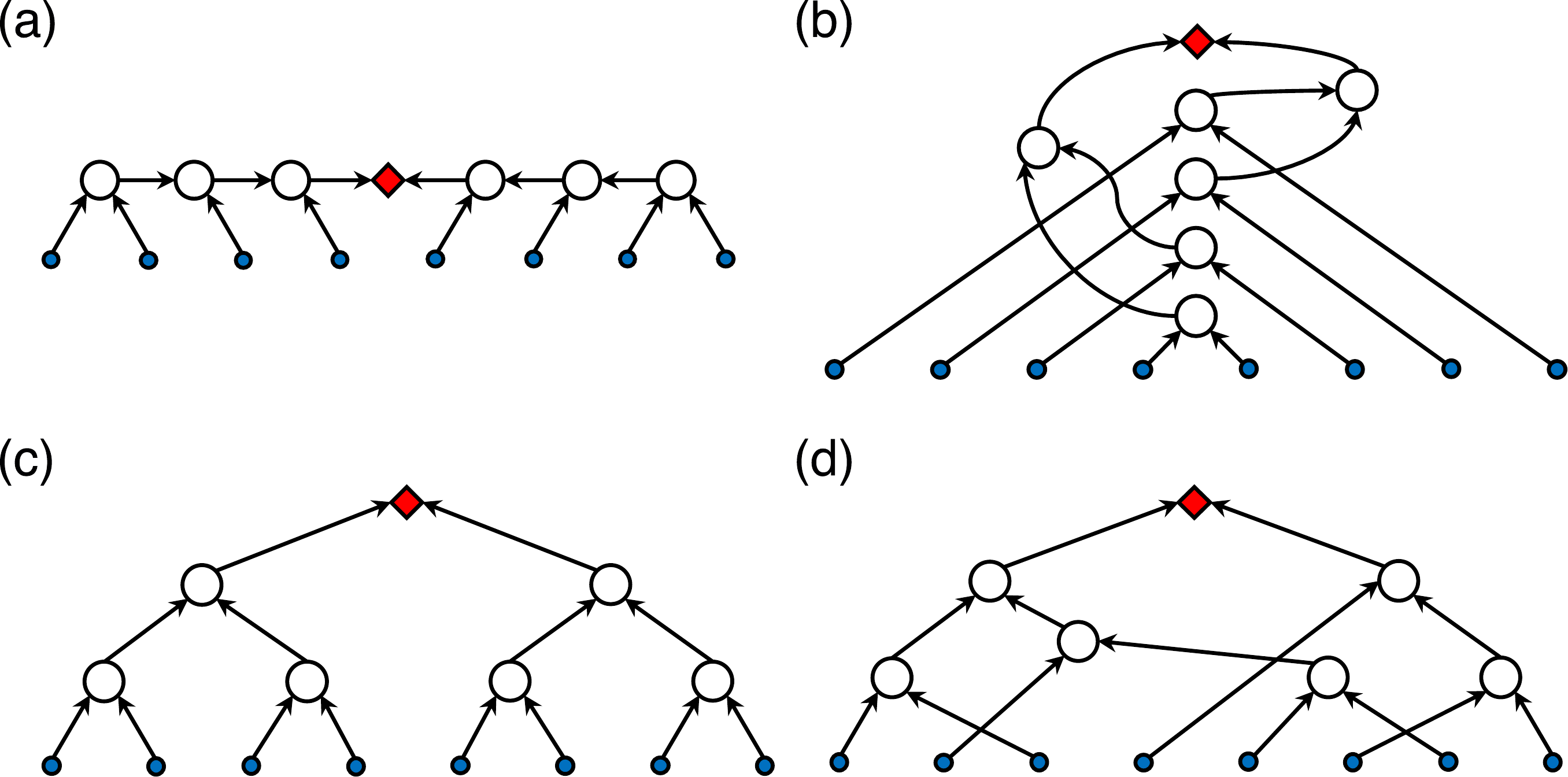}
  \caption{
  Examples of TTN structures: (a) a matrix product network (MPN) or tensor train, (b) a rainbow structural network, (c) a perfect binary tree (PBT) network, and (d) a general TTN.
  White circles represent tensors defined in Eq.~\eqref{eq:isometric condition}, while blue ones represent bare sites.
  Bare sites are arranged from left to right, following the order of basis states (indices).
  Arrows indicate tensor indices and point from the bare sites to the canonical center.
  These directions correspond to the domain and codomain of the isometric mapping as Eq.~\eqref{eq:isometric condition}.
  The red square highlights the singular value tensor at the canonical center.
  Notably, the position of the singular value tensor can be arbitrary under gauge transformations~\cite{PhysRevLett.91.147902, PhysRevLett.93.040502}.
}
\label{fig:variousTTN}
\end{figure}

In this study, we focus on the tree-tensor networks (TTNs)~\cite{PhysRevB.82.205105,larssonComputingVibrationalEigenstates2019}, which have no loop in their network structure; see Fig.~\ref{fig:variousTTN} for the examples.
The tree structure naturally allows us to impose the isometric conditions on tensors, resulting in efficient contraction schemes~\cite{PhysRevLett.99.220405,PhysRevB.79.144108, PhysRevLett.124.037201}.
In particular, the isometric conditions guarantee that a TTN can be brought into the form of Schmidt decomposition across any bipartition regions $A$ and $B$:
\begin{equation}\label{eq:schmidt_decomposition}
\ket{\Psi} = \sum_{p,q}\sum_c U_{pc}D_{c}V_{cq}\ket{\psi_p}^A\ket{\psi_q}^B~,
\end{equation}
where $U$ and $V$ are unitaries, and $D$ is the singular values tensor.
Here, TTN state $\ket{\Psi}$ belongs to the Hilbert space $\mathcal{H} = \mathcal{H}_A \otimes \mathcal{H}_B$, where $\{\ket{\psi_p}^A\}$ and $\{\ket{\psi_q}^B\}$ are orthonormal bases of the Hilbert spaces $\mathcal{H}_A$ and $\mathcal{H}_B$, respectively.
This representation makes it possible to take the truncation of $D_c$.
Additionally, the entanglement entropy (EE) of $\ket{\Psi}$ between $A$ and $B$ can be calculated as
\begin{align}\label{eq:entanglement}
\mathcal{S}&=- \operatorname{Tr}\left[\rho_{\mathrm{A}}\log\left(\rho_{\mathrm{A}}\right)\right] = -\operatorname{Tr}\left[\rho_{\mathrm{B}}\log\left(\rho_{\mathrm{B}}\right)\right] \nonumber \\
&= - \sum_{c}(D_c)^2\log(D_c)^2~,
\end{align}
where $\rho_A = \operatorname{Tr}_{B}(\ket{\Psi}\bra{\Psi})$ and $\rho_B = \operatorname{Tr}_{A}(\ket{\Psi}\bra{\Psi})$.

While introducing truncation of $D_c$ by bounding the dimension of tensors with $c \in [1, \chi]$ on Eq.~\eqref{eq:schmidt_decomposition} reduces computational complexity, it also leads to a loss of precision in the TTN representation.
Therefore, mitigating the loss of precision due to the finite bond dimension $\chi$ is a critical issue in the TN approach.

\begin{figure}[t]
  \centering
  \includegraphics[clip, width=3.3in]{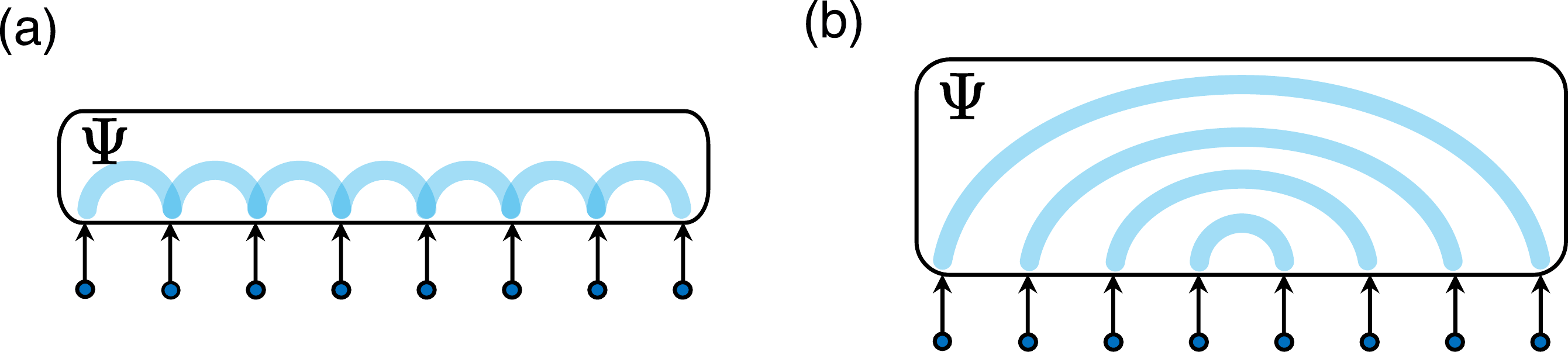}
  \caption{
  Examples of the entanglement structure of wavefunctions $\Psi$:
  (a) a state with a one-dimensional entanglement structure, and (b) a state with a rainbow bell pairs structure.
  }
  \label{fig:optimalTTN}
\end{figure}

A promising solution to the problem is to optimize the network structure.
The following examples illustrate the relevance of the network structure to the accuracy of the TTN approach.
Let us consider a quantum state with a one-dimensional (1D) entanglement pattern, where the entanglement between qubits arranged in 1D is short-ranged [Fig.\ \ref{fig:optimalTTN} (a)].
For this state, a Matrix Product Network (MPN)~\cite{PhysRevLett.123.170504,10.5555/2011832.2011833}, depicted in Fig.~\ref{fig:variousTTN} (a), also known as a tensor train~\cite{doi:10.1137/090752286}, is a reasonable choice.
As a result, MPN-based approaches, such as the density-matrix renormalization group (DMRG) method~\cite{whiteDensityMatrixFormulation1992,whiteDensitymatrixAlgorithmsQuantum1993,schollwockDensitymatrixRenormalizationGroup2011}, have been working well for 1D quantum systems.
As another example, consider a state where Bell-paired qubits are arranged in a rainbow pattern, as shown in Fig.~\ref{fig:optimalTTN}(b).
If the MPN is applied to represent this state accurately, the canonical center in Fig.~\ref{fig:variousTTN}(a) must accommodate an exponentially large dimension $\chi$ with respect to $N$, in order to carry an amount of entanglement equivalent to $N/2$ Bell pairs.
The bonds with a constant dimension $\chi$ miss such a large amount of entanglement, resulting in a significant loss of accuracy.
This problem can be resolved by employing a TTN with the structure shown in Fig.~\ref{fig:variousTTN}(b).
Using the TTN with this appropriate structure reduces the entanglement carried by each bond to that of, at most, a single Bell pair, allowing the state to be represented accurately with $\chi=2$.

As demonstrated by the above examples, the network structure can significantly affect the efficiency of the TTN approach~\cite{10.1093/ptep/ptad018}.
Nevertheless, identifying the optimal network structure remains a nontrivial problem, particularly for states with complex entanglement distributions.
Several studies have been conducted to develop methods for determining optimal TTN structures.
Many of these works have focused on optimizing the ordering of qudits in MPNs~\cite{chanHighlyCorrelatedCalculations2002,PhysRevB.68.195116,moritzConvergenceBehaviorDensitymatrix2005,PhysRevC.92.051303,liFlySwappingAlgorithm2022}, while others have attempted to explore optimal structures within TTNs~\cite{larssonComputingVibrationalEigenstates2019}.

An algorithm for searching the optimal TTN structure in variational calculations of quantum many-body systems has been proposed~\cite{hikiharaAutomaticStructuralOptimization2023}.
This algorithm focuses on a particular bond in the TTN and recombines the local network structure to minimize the entanglement brought by that bond.
Iterating this procedure while sweeping the entire network, the algorithm explores the TTN with the optimized structure.
This algorithm has been proven effective for several quantum spin models~\cite{hikiharaVisualizationEntanglementGeometry2024,hikiharaImprovingAccuracyTreetensor2025}.

In this work, we provide a software library for TTN calculations that includes the optimization of network structures.
The library consists of three main packages:
\begin{itemize}
\item[(1)] To perform the variational calculation for the lowest energy state of quantum spin systems.
TTNOpt provides the variational wave function with the optimized structure in the TTN format.
It outputs the optimized TTN, the variational energy, EE, and truncation error for each bond in the TTN, as well as the expectation values of various one- and two-spin functions.
TTNOpt can treat a wide range of quantum spin models. 
Namely, the Hamiltonian terms that can be handled include the XXZ/XYZ exchange, Dzyaloshinskii-Moriya, and symmetric off-diagonal exchange interactions for arbitrary spin pairs, as well as external magnetic fields and single-ion anisotropy for arbitrary spins. 
The interaction parameters can take different values depending on pairs or spins. 
The spin sizes can also be site-dependent. If the model treated has the U(1) symmetry of conservation of total magnetization, the user can specify the total magnetization of the subspace in which the variational calculation is performed.
\item[(2)]
To factorize or decompose a given high-dimensional tensor into a TTN with the optimized structure.
The user can input the tensor as a multidimensional function.
TTNOpt first decomposes the input tensor into the MPN and then performs the TTN structural optimization while maximizing the fidelity with the input tensor.
\item[(3)]
To reconstruct the network of a given TTN.
The user can input a tensor represented in the TTN format.
TTNOpt performs structural reconnection of the network for the TTN from the original input as the initial TTN.
While the TTN structure is optimized through sweeps, the tensors are only updated by the singular value decomposition (SVD).
\end{itemize}
Additionally, TTNOpt contains descriptions of sample calculations to demonstrate its functionality and practical applications.

The structure of this paper is as follows.
In Sec.~\ref{sec:usage}, we give an overview of the methods provided by TTNOpt and their basic usage.
We also describe the variables used for the calculation in detail.
In Sec.~\ref{sec:implemented algorithms}, we explain the implemented algorithms in TTNOpt.
While the explanation focuses on the sweep procedure proposed in Ref.~\cite{hikiharaAutomaticStructuralOptimization2023}, we also discuss the numerical techniques for TTN manipulations.
Then, we showcase example benchmarks for each implemented method in Sec.~\ref{sec:benchmark}.
Finally, Sec. \ref{sec:summary}
concludes with an outlook on TTNOpt and explores prospective avenues for future development.

\section{Basic usage of TTNOpt}
\label{sec:usage}
Here, we provide a detailed explanation of how to use TTNOpt.
This section describes the input files required mainly for ground-state searches and high-rank tensor factorizations.
The sample input files for the demonstrations discussed in Sec .~\ref {sec:benchmark}, including those used for the network reconstruction, are available in the ``sample'' directory of the GitHub repository~\cite{TTNOpt}.

\subsection{Ground state search}\label{subsec:usage gss}
The TTNOpt package has been developed for finite-size spin systems, including XXZ and XYZ Hamiltonians,
\begin{align}
\label{eq:xxz_hamiltonian}
H_{\text{XXZ}} &= \sum_{i,j(>i)}  J_{ij}\left( s_i^x s_j^x + s_i^y s_j^y + \Delta^z_{ij} s_i^z s_j^z \right )~, \\
\label{eq:xyz_hamiltonian}
H_{\text{XYZ}} &= \sum_{i,j(>i)}  \left(J^x_{ij} s_i^x s_j^x + J^y_{ij} s_i^y s_j^y + J_{ij}^z  s_i^z s_j^z\right)~, 
\end{align}
where $\bm s_i =(s^x_i,s^y_i,s^z_i)$ is a spin operator at $i$ th site on an $N$-site system, and $\{J_{ij}, \Delta^z_{ij}\}$ or $\{ J^x_{ij}, J^y_{ij}, J^z_{ij} \}$ are the coupling parameters for the exchange interactions between the $i$ th and $j$ th spins.
The size of spin ${\bm s}_i$ can be arbitrary for each site.
In addition to the Hamiltonian Eqs.~\eqref{eq:xxz_hamiltonian} and \eqref{eq:xyz_hamiltonian}, TTNOpt can treat the interactions including
magnetic field
\begin{equation}
\label{eq:magnetic field}
H_{{\rm{h}}_{\alpha}} = \sum_{i} - {\rm{h}}^{\alpha}_i s^\alpha_i~,
\end{equation}
single-ion anisotropy
\begin{equation}
\label{eq:single-ion anisotropy}
H_{\rm{D}} = \sum_{i}  {\rm{D}}_i (s^z_i)^2 ~,
\end{equation}
Dzyaloshinskii-Moriya (DM) interaction
\begin{equation}
\label{eq:DM interaction}
H_{{\rm{DM}}_{\alpha}} = \sum_{i,j(>i)}  {\rm{D}}^{\alpha}_{ij} (\bm{s}_i \times \bm{s}_j )^{\alpha} ~,
\end{equation}
and symmetric off-diagonal exchange anisotropy
\begin{equation}
\label{eq:gamma interaction}
H_{{\Gamma}_{\alpha}} = \sum_{i,j(>i)} \Gamma^{\alpha}_{ij}(s_i^\zeta s_j^\eta + s_i^\eta s_j^\zeta) ~,
\end{equation}
where $\alpha, \zeta, \eta \in \{x,y,z\}$ and $\alpha \neq \zeta \neq \eta$.

Regarding the whole Hamiltonian $H = H_{\text{XXZ}} + \sum_{\alpha\in \{ x, y, z\}}H_{{\rm{h}}_{\alpha}} + H_{\rm{D}} + \sum_{\alpha\in \{ x, y, z\}}H_{{\rm{DM}}_{\alpha}} + \sum_{\alpha\in \{ x, y, z\}}H_{{\Gamma}_{\alpha}}$, if and only if it meets ${\rm{h}}^x_i = {\rm{h}}^y_i = {\rm{D}}^x_{ij} = {\rm{D}}^y_{ij} = {\Gamma}^{x}_{ij} ={\Gamma}^{y}_{ij} = {\Gamma}^{z}_{ij} = 0$, the Hamiltonian commutes with the operator for the total magnetization of $z$ axis:
\begin{equation}
    M=\sum_i{ s^z_i}~,
\end{equation}
i.e., $[H, M] = 0$, so that the U(1) symmetry is preserved.
In this case, TTNOpt provides a function to calculate the TTNs for the lowest-energy state within the subspace labeled by $M$.

\subsubsection{How to set input files}\label{subusubsec:input gss}
Running the TTNOpt package requires the main input file and, if necessary, several setting files.
The main input file consists of {\bf system}, {\bf numerics}, and {\bf output}.
The meaning of each section and the variables used there are explained below.

\begin{description}
\item{\underline{system}}

This section requires users to specify the information of the Hamiltonian for the target system, including the number of spins, spin size, and interactions in the Hamiltonian.
TTNOpt requires users to prepare a separate file to define the two-site interactions of Eq.~\eqref{eq:xxz_hamiltonian} or Eq.~\eqref{eq:xyz_hamiltonian}.
Other additional terms Eqs.~\eqref{eq:magnetic field}, ~\eqref{eq:single-ion anisotropy}, ~\eqref{eq:DM interaction}, and \eqref{eq:gamma interaction} are not necessarily specified if they are not present in the Hamiltonian.
TTNOpt assumes that the input files are in the ``.dat'' format.

\begin{itemize}
\item {\bf N} (INTEGER): \\
The number of spins, $N$.
Thus, TTN wave functions are tensors in the vector product space of $N$ local spin Hilbert spaces.

\item {\bf spin\_size} (REAL or STRING) \\
The spin sizes $s_i$ defined for $i \in [0, N-1]$.  
If a spin value $s$ is provided, it is applied uniformly across the entire system, i.e., $s_i = s$ for all $i \in [0, N-1]$.  
Alternatively, if a file path is provided, TTNOpt imports the file, which must contain two columns:  
The first column specifies the site index $i$, and the second column specifies the corresponding spin value $s_i$.
Note that, in the case that the spin size is a half-odd integer, TTNOpt does not accept decimal values for spin settings, and only the fractional representations, such as $1/2$ and $3/2$, are allowed.

\item {\bf model.type} (XXZ or XYZ) \\
The basic interaction type, XXZ or XYZ interaction, that is respectively given in Eq.~\eqref{eq:xxz_hamiltonian} or \eqref{eq:xyz_hamiltonian}.
\item {\bf model.file} (STRING) \\
The coupling parameters for the exchange interactions.
Each row contains two integers and two or three floats, where the first two columns specify a pair of site indices $i, j$ and subsequent columns specify the coupling parameters $J_{ij}, \Delta^z_{ij}$ or $J^x_{ij}, J^y_{ij}, J^z_{ij}$ according to {\bf system.model.type}.
\item {\bf MF\_X~(Y,~Z)} (REAL or STRING) \\
The magnetic field along the $\alpha$ direction $(\alpha=x,y,z)$ described in Eq.~\eqref{eq:magnetic field}.
If a float value $\rm{h}$ is provided, it is applied uniformly across the entire system, i.e., ${\rm{h}}^{\alpha}_i = \rm{h}$ for $i \in [0, N-1]$.
Alternatively, if the path to a file is provided, TTNOpt imports the file, which must contain two columns where the first and second columns specify, respectively, the site $i$ and the corresponding value ${\rm{h}}^{\alpha}_i$.

\item {\bf SIA} (REAL or STRING) \\
The single-ion anisotropy described in Eq.~\eqref{eq:single-ion anisotropy}.
If a float value ${\rm{D}}$ is provided, the anisotropy is applied uniformly across the entire system, i.e., ${\rm{D}}_i = {\rm{D}}$ for $i \in [0, N-1]$.
Alternatively, if the path to a file is provided, TTNOpt imports the file, which must contain two columns where the first and second columns specify, respectively, the site $i$ and the corresponding value ${\rm{D}}_i$.

\item {\bf DM\_X~(Y,~Z)} (STRING) \\
The Dzyaloshinskii-Moriya (DM) interaction described in Eq.~\eqref{eq:DM interaction}.
TTNOpt requires a file with three columns to contain two integers and a real value.
The first two columns identify a pair of site indices $i, j$, and the last column specifies ${\rm{D}}^{\alpha}_{ij}$.

\item {\bf SOD\_X~(Y,~Z)}(STRING) \\
The symmetric off-diagonal anisotropic exchange interaction described in Eq.~\eqref{eq:gamma interaction}.
TTNOpt requires a file with three columns to contain two integers and a real value.
The first two columns identify a pair of site indices $i, j$, and the last column specifies $\Gamma^{\alpha}_{ij}$.

\end{itemize}
\end{description}

\begin{description}
\item{\underline{numerics}}

This section requires users to specify the conditions and hyperparameters for the calculation, including the settings for the structural optimization algorithm and the maximum bond dimension of the TTN states.

\begin{itemize}
\item {\bf init\_tree} (0 or 1) \\
If this value is set to $0$, the initial structure is set to the MPN.
If it is set to $1$ and the system size is a power of 2, the initial structure is set to the perfect binary tree (PBT) structure.
Otherwise, the MPN structure is used by default.

\item {\bf initial\_bond\_dimension} (INTEGER) \\
The maximum bond dimension $\chi_{\rm{init}}$ during the preparation of an initial TTN.
The initial tensors are prepared by the real space renormalization group (RSRG)~\cite{wilsonRenormalizationGroupCritical1975,hikiharaNumericalRenormalizationgroupStudy1999}, where the bond dimension of tensors is upper-bounded by $\chi_{\rm init}$~(see Sec.~\ref{subsec:initializing tensors} for details). The value of $\chi_{\rm init}$ must not be too small to ensure that each tensor contains all degenerate lowest-energy states of the block Hamiltonian [Eq.~\eqref{eq:block Hamiltonian} in Sec.~\ref{sec:representation of the Hamiltonian}] in the renormalized region belonging to the tensor.

\item {\bf total\_magnetization} (REAL) \\
The total magnetization $M$ that defines the subspace in which the lowest-energy state is computed.
This input is used only for the Hamiltonian with U(1) symmetry.
To activate this function, it is not allowed to use XYZ interaction in {\bf{system.model.type}} even if the user numerically sets $J_x = J_y$.
Among other terms, those that break the symmetry are also prohibited.

When the initial bond dimension $\chi_{{\rm{init}}}$ is not sufficiently large, the RSRG method 
may fail to construct the initial TTN in the subspace labeled by the magnetization $M$.
In this case, TTNOpt initializes the TTN state with a magnetization $M'$ such that $|M'| < |M|$ and $|M'|$ is the closest to $|M|$ among the magnetizations that can be spanned by the RSRG.
It then performs warmup calculations while keeping the TTN state's bond dimension at $\chi_{\rm{init}}$ and adjusting $M'$ as $M':= M' \pm 1$ at the end of each update sweep~\cite{note3}, until the appropriate initial state with magnetization $M$ is realized.

\item {\bf opt\_structure.type} (0, 1 or 2) \\
The method for optimizing the network structure.
If the value is $0$, TTNOpt does not optimize the TTN structure.
If it is $1$, the TTNOpt performs structural reconstruction by referring to EEs.
If the value is $2$, TTNOpt selects the structure with the minimum truncation error [Eq.~\eqref{eq:truncated singular values} in Sec.~\ref{subsec:main procedure}].
However, when the differences between the minimum truncation errors are less than $1 \times 10^{-13}$, the optimal structure is determined using EE as a secondary criterion.

\item {\bf opt\_structure.temperature} (REAL) \\
An effective temperature $T_0$, which is a positive real number. When {\bf opt\_structure.type} = 1, the structure is selected stochastically using this $T_0$ with $n_{\tau}$, which is described later in Sec.~\ref{subsec:main procedure}.
This value is set to $T_0 = 0$ by default, and then TTNOpt chooses the structure with the minimum EE.

\item {\bf opt\_structure.tau} (INTEGER) \\
The decay factor $n_{\tau}$, which is a positive integer. When {\bf opt\_structure.type} = 1 and $T_0 > 0$, it controls the temperature decay according to the sweep count $n$. [See Eq.~\eqref{eq:manipulate temperature} in Sec.~\ref{subsec:main procedure}.]
TTNOpt sets $n_{\tau}$ to $\lfloor n_{\rm max, 0} /2 \rfloor$ by default.

\item {\bf opt\_structure.seed} (INTEGER) \\
The random seed for stochastic selection of the structure. [See Eq.~\eqref{eq:stochastic selection} in Sec.~\ref{subsec:main procedure}.]
TTNOpt sets this value to $0$ by default to ensure that the results are reproducible.

\item {\bf max\_bond\_dimensions} (LIST of INTEGER) \\
The elements of the list specify the maximum bond dimensions $\bm{\chi}=[\chi_m]_{1\leq m\leq\mathfrak {m}}$ in the TTN, where $\mathfrak{m}$ is the total number of stages.
At the $m$-th stage, TTNOpt performs sweeps using $\chi := \chi_m$ until the TTN state converges or the number of sweeps reaches $n_{\max, m}$, as specified in {\bf numerics.max\_num\_sweeps}.
Once the sweeps at stage $m$ are completed, TTNOpt proceeds to the next stage with $\chi := \chi_{m+1}$ and $n_{\max, m+1}$, using the TTN state obtained at $\chi = \chi_m$ as the initial state.
It is worth noting that the values in $\bm{\chi}$ should be arranged in ascending order.
This strategy enables computation with larger $\chi_m$, which requires a higher computational cost, to begin from a well-prepared initial state.
As a result, the number of sweeps needed for convergence at larger $\chi_m$ can be reduced.
If ${\bf{numerics.opt\_structure.type}} \in \{1, 2\}$, the structural optimization is applied {\it{only for}} the sweeps with $\chi = \chi_1$.
As for the remaining computation with $\chi=\chi_m$ for $m > 1$, only tensors are updated while the structure is fixed.

\item {\bf max\_num\_sweeps} (LIST of INTEGER) \\
The elements of the list specify the maximum number of sweeps $\bm{n}_{\max} = [ n_{\max,m}]_{1 \leq m \leq \mathfrak{m}}$ for each stage of calculations.
All elements $n_{\max, m}$ should be set sufficiently large to achieve the TTN state's convergence.
In this paper, we denote $n_{\max,m}$ simply as $n_{\max}$ to represent the maximum number of sweeps at $m$ th stage unless otherwise specified.

\item {\bf energy\_convergence\_threshold} (REAL) \\
The tolerance $\epsilon_E$ for the convergence of energy.
If the relative difference between each the energy ${\mathfrak{E}}_b$ calculated by using the Lanczos method for the auxiliary bond $b$ at the current sweep and that from the previous sweep, $\mathfrak{E}'_b$, is less than the threshold $\epsilon_{\rm{E}}$, i.e., $|1 - \frac{{\mathfrak{E}}'_b}{{\mathfrak{E}}_b}| < \epsilon_{\rm{E}}$ for all $b$, TTNOpt considers that the TTN has been converged concerning energy.
TTNOpt sets this value to  $\epsilon_{\rm{E}} = 1 \times 10^{-8}$ by default.

\item {\bf{entanglement\_convergence\_threshold}} (REAL) \\
The tolerance $\epsilon_S$ for the convergence of EE.
If the difference between the bipartite EE $\mathfrak{S}_b$ for the bond $b$ at the current sweep as Eq.~\eqref{eq:entanglement} and that from the previous sweep, $\mathfrak{S}'_b$, is less than the threshold $\epsilon_{\mathcal{S}}$, i.e., $|\mathfrak{S}_b - \mathfrak{S}'_b| < \epsilon_{\mathcal{S}}$ for all bonds $b$, TTNOpt considers that the EEs of TTN have been converged.
Furthermore, if {\bf{opt\_structure.type}} is set to $1$ and the EE of the optimal and previous structures differs by less than $\epsilon_{\mathcal{S}}$, the structure remains unchanged to avoid inconsequential fluctuation of the TTN structure at each step of sweeps.
TTNOpt sets this value to  $\epsilon_{\mathcal{S}} = 1 \times 10^{-8}$ by default.

\item {\bf energy\_degeneracy\_threshold} (REAL) \\
The threshold $\delta_{{\rm{E}}}$ used in the preparation of the initial tensors by the RSRG to determine whether the eigenvalues of a block Hamiltonian are degenerate.
The procedure is detailed in Sec.~\ref {subsec:initializing tensors}.
TTNOpt sets this value to  $\delta_{\rm{E}} = 1 \times 10^{-8}$ by default.

\item {\bf entanglement\_degeneracy\_threshold} (REAL) \\
The threshold $\delta_{\mathcal{S}}$ used in the SVD to determine whether singular values are degenerate for updating the local two-tensor.
The procedure is detailed in the last paragraph of Sec.~\ref{subsec:initializing tensors}.
TTNOpt sets this value to  $\delta_{\mathcal{S}} = 1 \times 10^{-8}$ by default.
\end{itemize}
\end{description}

\begin{description}
\item{\underline{output}}

This section requires users to specify the physical quantities for which TTNOpt will generate output files.
By default, TTNOpt outputs, in a file named ``basic.csv'', the EEs for all bonds, as well as the variational energies and truncation errors for the auxiliary bonds, computed during the final sweep of each stage $m \in [1, \mathfrak{m}]$.
In the ``basic.csv'' file, all bonds are identified by two nodes connected by the bond, $(i, i')$ with $i, i' \in [0, N+N_{\rm t}-1]$, where $N$ is the number of spins and $N_{\rm t}$ is the number of tensors.
Additionally, TTNOpt saves the set of bond labels $\bm E$ for the TTN structure in a file named ``graph.dat''.
The format to describe the bond labels is explained in the second paragraph of Sec.~\ref{subsec:representation of TTN}.
Users can save single-site spin expectation values at each site and two-site spin correlations between any two sites.

\begin{itemize}

\item {\bf dir} (STRING) \\
The location of the directory where the data will be output.

\item {\bf single\_site} (0 or 1) \\
If this value equals $1$, TTNOpt calculates single-site spin expectation values and saves them in a file named ``single\_site.csv''.
The file has four columns where the first column owns site $i$ with $i \in [0, N-1]$ and the subsequent columns have $\langle s^{\alpha}_i
\rangle$ with $\alpha = x, y, z$ , where $\langle \cdots \rangle$ denotes the expectation value of $\cdots$ in the ground state.
Otherwise, TTNOpt does not save them.
\item {\bf two\_site} (0 or 1) \\
If this value equals $1$, TTNOpt calculates two-site spin correlation functions and saves them in a file named ``two\_site.csv''.
The file has eleven columns where the first two columns own a pair of two sites $i, j$ with $i, j \in [0, N-1]$ and $i < j$, and the subsequent columns have $\langle s^{\alpha}_is^{\beta}_{j} \rangle$, where $(\alpha, \beta) \in \{ xx,yy,zz,yz,zy,zx,xz,xy,yx \}$.
Otherwise, TTNOpt does not save them.

\end{itemize}
\end{description}

\subsubsection{Run and results}
After preparing all input files described above, users can perform the calculation as follows:
\begin{quote}
\begin{itemize}
\$ gss input.yml
\end{itemize}
\end{quote}
Here, the results of $m$ th stage of calculation are saved in a subdirectory ``run$\{m\}$'', where $\{m\}$ with $m \in [1, \mathfrak{m}]$ relies on variable expansion of Python notation, under the directory specified by the {\bf{output.dir}} in the main input file.

\subsection{Factorising tensors}\label{subsec:usage ft}
\begin{figure}[t]
  \centering
  \includegraphics[clip, width=3.3in]{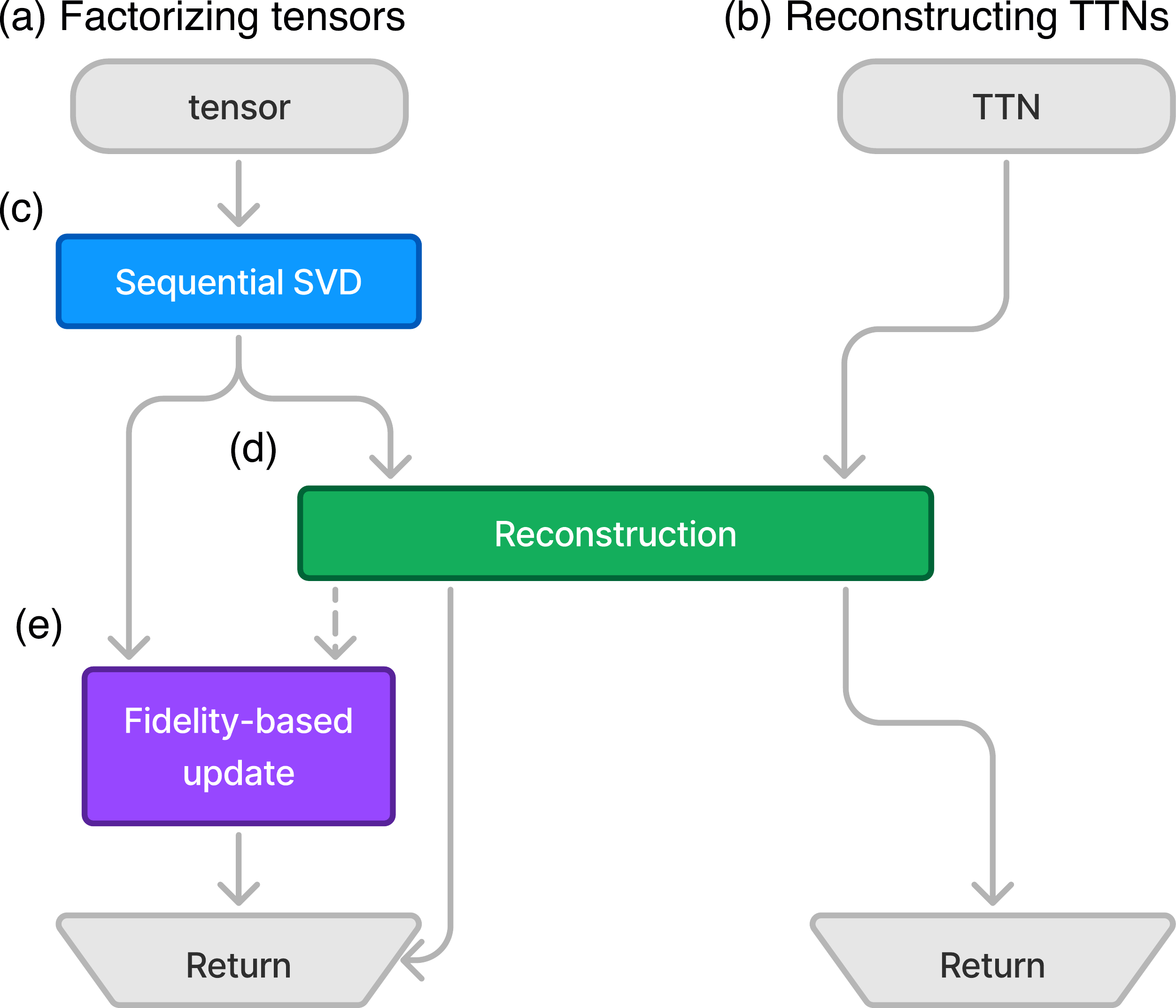}
  \caption{
  Schematic overview of the procedures for (a) factorizing tensors and (b) reconstructing TTNs.
  In (a), TTNOpt performs (c) sequential SVD to construct an MPN, after which users might either (d) apply reconstruction to this MPN or proceed directly to (e), where both the tensors and the network structure of the TTN are optimized to maximize the fidelity with $\Psi$.
  TTNOpt also allows using the optimized TTN obtained in (d) as the input to (e), depicted as a dashed line.
  In (b), TTNOpt runs (d) for the given TTN.
  }
  \label{fig:ft_commands}
\end{figure}

The TTNOpt package provides functions to factorize a high-dimensional tensor $\Psi=\{\Psi_{s_0, \dotsc, s_{N-1}}\}$, with indices $s_0, \dotsc, s_{N-1}$, into a TTN as an efficient data structure.  
The overall procedure is illustrated in Fig.~\ref{fig:ft_commands}.

TTNOpt first decomposes the input tensor $\Psi$ into an MPN using sequential SVD [Fig.~\ref{fig:ft_commands}(c)]~\cite{schollwockDensitymatrixRenormalizationGroup2011} .  
Users may then choose to transform this MPN into a TTN via reconstruction sweeps [Fig.~\ref{fig:ft_commands}(d)].  
This yields a TTN structure optimized to minimize bipartite EEs based on local SVDs applied to the initial MPN.

If the entanglement structure of the input tensor $\Psi$ is not compatible with the MPN, the TTN obtained from this procedure may suffer from low-precision approximations.  
To address this limitation, TTNOpt also implements the fidelity-based update that directly references $\Psi$ [Fig.~\ref{fig:ft_commands}(e)].  
Specifically, this procedure optimizes the tensor elements to maximize the fidelity with 
$\Psi$, while simultaneously reconstructing the network structure to reduce EEs.
However, since it requires explicit contractions with $\Psi$ at each update step, it entails a significantly higher computational cost compared to the reconstruction process alone.

\subsubsection{How to set input files}\label{sec:input file of ft}
\begin{description}
\item{\underline{target}}

This section requires specifying the directory of the input tensor $\Psi$.
TTNOpt needs users to specify ``.npy'' format file as tensors.

\begin{itemize}
\item {\bf tensor} (STRING) \\
The directory name for the file of the input tensor $\Psi$.
\end{itemize}
\end{description}

\begin{description}
\item{\underline{numerics}}

This section requires users to specify the conditions for the calculation.

\begin{itemize}
\item {\bf initial\_bond\_dimension} (INTEGER) \\
The initial bond dimension $\chi_{\rm{init}}$.
TTNOpt first decomposes $\Psi$ into an MPN structure with up to this bond dimension $\chi_{\rm{init}}$ using the SVD.

\item {\bf opt\_structure.type} (0, 1 or 2) \\
This value is applied to local reconstructions in TTN structural optimization.
Explantions of this input and the following ones, {\bf opt\_structure.temperature}, {\bf opt\_structure.tau}, and {\bf opt\_structure.seed} are in Sec.~\ref{subusubsec:input gss}.

\item {\bf max\_sweep\_num} (INTEGER) \\
The maximum number of sweeps $n_{\max}$ for reconstruction.
During the sweep procedure, we set the bond dimension $\chi$ as $\chi_{\rm{init}}$ specified by {\bf{initial\_bond\_dimension}}.

\item {\bf entanglement\_convergence\_threshold} (REAL) \\
The entanglement convergence threshold $\epsilon_{\mathcal{S}}$, which is explained in Sec.~\ref{subusubsec:input gss}.
It is noted that this value is used both when TTNOpt reconstructs the initial MPN into the optimal TTN and updates the TTN state based on the fidelity with the input tensor $\Psi$.

\item {\bf max\_truncated\_singularvalue} (REAL) \\
This threshold $\sigma$ is used in SVD to reduce the bond dimension while tolerating a certain loss of accuracy.
All singular values satisfying $D_i / D_1 \leq \sigma$ are truncated, where $D_i$ is the $i$-th singular value sorted in descending order.
By default, $\sigma$ is set to 0, in which case up to $\chi$ singular values are retained.

\item {\bf fidelity.opt\_structure.type} (0, 1 or 2) \\
This value is applied when TTNOpt updates the TTN state according to the fidelity.
Explantions of this value and the following inputs, {\bf fidelity.opt\_structure.temperature}, {\bf fidelity.opt\_structure.tau}, {\bf fidelity.opt\_structure.seed}, and {\bf fidelity.max\_num\_sweeps} are in Sec.~\ref{subusubsec:input gss}.
Note that if users set values related to the fidelity-based optimization in the input file, TTNOpt will update both the network structure and tensors of the TTN concerning $\Psi$, even if the TTN structure has already been determined by {\bf opt\_structure.type}.

\item {\bf fidelity.max\_bond\_dimensions} (LIST of INTEGER) \\
Explanations of this value are detailed in Sec.~\ref {subusubsec:input gss}. 
The bond dimension practically kept can be reduced when a nonzero $\sigma$ is set in {\bf max\_trunated\_value}.

\item {\bf fidelity.convergence\_threshold}(REAL) \\
The tolerance $\epsilon_F$ for the convergence of fidelity.
If the difference between the fidelity $F_b$ calculated for auxiliary bond $b$ at the current sweep and that from the previous sweep, $F'_b$, is less than the threshold $\epsilon_{F}$, i.e., $|F_b - F'_b| < \epsilon_{F}$ for all auxiliary bonds $b$, TTNOpt considers that the TTN has been converged concerning fidelity.

\end{itemize}
\end{description}

\begin{description}
\item{\underline{output}}

This section requires users to specify the output settings.
By default, TTNOpt outputs properties of TTN in a file named ``basic.csv'' and TTN structure in ``graph.dat'' in the same manner as the ground state search algorithm.
In the ``basic.csv'', TTNOpt outputs the EEs for all bonds and the truncation errors for all auxiliary bonds.
TTNOpt also saves the fidelity with the input tensor $\Psi$ calculated at each auxiliary bond to a file named ``basic.csv'', if users specify the input file to conduct the update regarding the fidelity.

In addition to the outputs above, the user can save the tensors, singular values, and the norm of the tensor $\Psi$, which is used for the normalization of the TTN.

\begin{itemize}
\item {\bf dir} (STRING) \\
The location of the directory where the data will be output.
\item {\bf tensors} (0 or 1) \\
If this value is 1, TTNOpt saves optimized tensors as the ``isometry$\{i\}$.npy'' file with $i \in [0, N_{\rm t} 
- 1]$, singular values tensor in the ``singular\_values.npy'' file, and norm in the ``norm.npy'' file.
Otherwise, TTNOpt does not save them.
\end{itemize}
\end{description}

\subsubsection{Run and result}
After preparing the input files described above, users can perform the calculation as follows:
\begin{quote}
\begin{itemize}
\$ ft input.yml
\end{itemize}
\end{quote}
Here, the computed results are output in the directory specified by the {\bf output.dir} variable in the main input file.

\subsection{Reconstruction of TTNs}

Motivated by the recent applications of MPNs and tensor train data, TTNOpt allows users to load a TTN.
TTNOpt then reconstructs the network structure of a given TTN [Figs.~\ref{fig:ft_commands}(b) and (d)].

This function of TTNopt would be powerful for searching more efficient TTN structures in the sense that each bond carries low EE or has small bond dimensions.

Users can perform the calculation as follows:
\begin{quote}
\begin{itemize}
\$ ft input.yml
\end{itemize}
\end{quote}
Note that the command to execute this method is identical to the one used for the factorizing tensors.
The users must ensure that the input tensor is consistent with the intended type, which is either a tensor or a TTN (see the sample input files in Ref.~\cite{TTNOpt}).
The variables for the calculations share the same format as those for the factorizing tensors method, excluding those relevant to the fidelity-based optimization, described in Sec.~\ref{sec:input file of ft}.

\section{Inplemented algorithms}\label{sec:implemented algorithms}
\subsection{Representation of TTNs}\label{subsec:representation of TTN}
The TTNOpt package constructs TTN states from a set of three-leg isometric tensors $\bm{v}=\{v_0,v_1,\dotsc,v_{N_{\rm t}-1}\}$, where $N_{\rm t}=N-2$ is the total number of tensors
~\cite{note1}. 
Each isometric tensor $\big[v_i\big]^{i_3}_{i_1i_2}$ for all $0 \leq i \leq N_{\rm t}-1$, where $i_k$ with $k\in\{1,2,3\}$ represents index with bond dimensions $\chi_{i_k}$, satisfies the following isometric condition:
\begin{equation}\label{eq:isometric condition}
\sum_{i_1i_2}\big[v_i \big]^{i_3}_{i_1i_2}\big[v^{*}_i \big]^{i_3'}_{i_1i_2}=\delta_{i_3i_3'}~,
\end{equation}
where $v^{*}_i$ denotes the complex conjugate of $v_i$, and $\delta_{i_3i_3'}$ is the Kronecker delta.
On the isometry $\big[v_i\big]^{i_3}_{i_1i_2}$, the degrees of freedom $\chi_{i_3}$ is limited by the product of the bond dimensions $\chi_{i_1}$ and $\chi_{i_2}$, i.e., $\chi_{i_3} \leq \chi_{i_1}\chi_{i_2}$. 
In addition, TTNOpt practically poses an upper bound $\chi$ on bond dimensions for all bonds in the TTN state.

\begin{figure}[t]
  \centering
  \includegraphics[clip, width=3.3in]{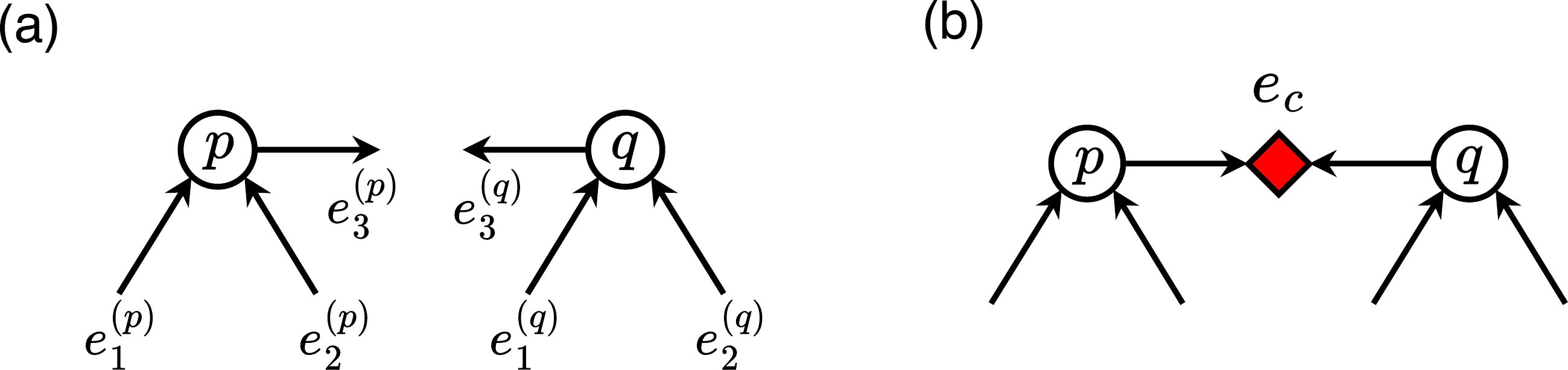}
  \caption{Schematic diagrams of 
  (a) data structure of labels for tensors and edges around the initial canonical center, i.e., $e^{(p)}_3 = e^{(q)}_3$, and
  (b) local structure surrounding the canonical center $e_{\rm c}$.
  }
  \label{fig:canonical_center}
\end{figure}

TTNOpt maintains connectivity between tensors as a set of three bond (edge) labels $\bm{E} = \{E_0, \dotsc, E_{N_{\rm t} - 1} \}$ where $E_{i} = (e^{(i)}_1, e^{(i)}_2, e^{(i)}_3)$, to define TTN structures.
Each edge label $e$ is written as an integer $e \in [0, 2N_{\rm t}]$ where $2N_{\rm t}+1$ is the number of bonds in the TTN.
The edge with the label $e = r \in[0, N-1]$ is connected to the bare spin $s_{r}$.
In this data structure, two tensors $v_i$ and $v_j$ that share the same edge label in $E_i$ and $E_j$ are connected.
There is an unique pair of tensors specified by $p$ and $q$ connected to each other through $e^{(p)}_3$ and $e^{(q)}_3$ as shown in Fig.~\ref{fig:canonical_center}(a).
This bond is referred to as the canonical center $e_{\rm c}$, defined as $e_{\rm c} = e^{(p)}_3 = e^{(q)}_3$ detailed in Fig.~\ref{fig:canonical_center}(a) and (b).
It is important to note that in TTNOpt, $e_{\rm c}$ is supposed not to be connected to any physical sites, i.e., $e_{\rm c} \notin [0, N-1]$.
TTNOpt assumes that up to $\chi$ elements from the vector $D = (D_1, \cdots, D_{\chi^2})$ are assigned on $e_{\rm c}$, where the singular values in $D$ are ordered in descending magnitude as $D_1 \ge D_2 \ge \cdots \ge D_{\chi^2} \ge 0$.
Since the truncation of bond dimensions causes a loss of the norm of TTN states, TTNOpt rescales the singular values, ${\mathfrak{D}_c}$ with $c\in[1, \chi]$ as
\begin{equation}\label{eq:normalize singular values}
    \mathfrak{D}_c := \frac{D_c}{\sqrt{\sum^{\chi}_{c'=1}(D_{c'})^2}}~.
\end{equation}
That is because, within the canonical formulas, the norm of the TTN state $\ket{\Psi}$ is described as $\langle \Psi | \Psi \rangle = \sum^{\chi}_{c'=1}{(D_{c'})^2}$.
With these settings, TTNOpt allows for representing TTN states in mixed canonical form~\cite{schollwockDensitymatrixRenormalizationGroup2011} to manage various calculations efficiently.

\subsection{Ground state search}

\subsubsection{Representation of the Hamiltonian}\label{sec:representation of the Hamiltonian}

Conducting the ground-state search requires constructing the effective Hamiltonian for each renormalized region, which is achieved by contracting a tensor network composed of the TTN state and the local Hamiltonian tensors.
TTNOpt preserves the renormalized spin operators, $\tilde{S}^z$ and $\tilde{S}^+$, at all bonds, enabling efficient evaluation of these contractions.
Since $\tilde{S}^-$ is the Hermitian conjugate of $\tilde{S}^+$, it is sufficient to retain only the $\tilde{S}^z$ and $\tilde{S}^+$ operators~\cite{whiteDensityMatrixFormulation1992, whiteDensitymatrixAlgorithmsQuantum1993}.

For a general TTN structure, this procedure differs from that in the case of considering only MPN, where the entire contraction with matrix product operators is efficient with $O(N\chi^3)$ 
 cost~\cite{schollwockDensitymatrixRenormalizationGroup2011}.

To manage the calculation on TTN, we assign a distance $d_{e^{(i)}_3} \in [0, N_{\rm t}-1]$ to each edge label $e^{(i)}_3$ for all $v_i$, measured from the origin bond $o_{\rm c}$, which corresponds to the canonical center of the initialized TTN and $d_{o_{\rm c}} = 0$.
Specifically, the value $d_{e^{(i)}_3}$ represents the minimum number of isometries that must be traversed to reach $v_i$ from $o_{\rm c}$.
We also introduce ${\bm L}=\{{\bm l}_{0},{\bm l}_1,\dots,{\bm l}_{{d}_{\rm max}} \}$ with ${\bm l}_{\mathfrak{d}} = \{ l^1_\mathfrak{d}, \dots, l^{n_\mathfrak{d}}_\mathfrak{d} \mid l_{\mathfrak{d}}^i \in [0, N_{\rm t} - 1] \}$, where $n_\mathfrak{d}$ is the number of edges whose distances are equal to $\mathfrak{d}$, i.e., $d_{ l^1_\mathfrak{d}} = d_{ l^2_\mathfrak{d}} = \dots = d_{ l^{n_\mathfrak{d}}_\mathfrak{d}} = \mathfrak{d}$).
Regarding the maximum distance $d_{\rm max}$, the isometries $\{ v_{i} \mid e^{(i)}_3 \in \bm{l}_{d_{\rm max}} \}$ are ensured to connect directly to bare sites, i.e., $e^{(1)}_i, e^{(2)}_i \in [0, N-1]$.
For example, in the MPN structure as Fig.~\ref{fig:variousTTN}(a), in which the origin bond $o_{\rm c}$ is positioned on the midpoint of the system, $\{ v_{i} \mid e^{(i)}_3 \in \bm{l}_{d_{\rm max}} \}$ correspond to two isometries located at both ends of the MPN.
Additionally, to define ${\bm L}$ uniquely, we incorporated an ordering rule such that $l^1_{\mathfrak{d}} < l^2_{\mathfrak{d}} < \cdots < l^{n_\mathfrak{d}}_{\mathfrak{d}}$ for each ${\bm l}_{\mathfrak{d}}$.
We finally introduce a set of spin locations ${\bm r}^{(i)}_k = \{ r^{(i)}_{k,1}, r^{(i)}_{k,2}, \cdots, r^{(i)}_{k,g(i,k)} \}$ within the renormalized region specified by the edge label $e^{(i)}_k$, where $g(i,k) = |{\bm r}^{(i)}_k|$.
It is trivial that ${\bm r}^{(i)}_3={\bm r}^{(i)}_1 \cup {\bm r}^{(i)}_2$, since the sets of spins of $e^{(i)}_1$ and $e^{(i)}_2$ are renormalized to $e^{(i)}_3$ by $v_i$.

These definitions allow us to consider the following renormalized spin transformation using the isometry $v_i$:
\begin{equation}\label{eq:block spin operator}
\left[ \tilde{S}^{(\cdot)}_{e^{(i)}_3,r} \right]_{i_3,i'_3} = 
\left\{
\begin{array}{ll}
\sum\limits_{i_1i_2i'_1} \Big[v_i\Big]^{i_3}_{i_1,i_2} \left[\tilde{S}_{e^{(i)}_1,r}^{(\cdot )}\right]_{i_1i'_1} \Big[v^*_i\Big]^{i'_3}_{i'_1,i_2} & (r \in {\bm r}^{(i)}_1)\\
\sum\limits_{i_1i_2i'_2} \Big[v_i\Big]^{i_3}_{i_1,i_2} \left[\tilde{S}_{e^{(i)}_2,r}^{(\cdot )}\right]_{i_2i'_2} \Big[v^*_i\Big]^{i'_3}_{i_1,i'_2} & (r \in {\bm r}^{(i)}_2)
\end{array}
\right. ~.
\end{equation}
If $e^{(i)}_k = r$, $\tilde{S}^{(\cdot)}$ is equal to the bare spin operator $s^{(\cdot)}$ where $(\cdot)\in \{z,+\}$ indicates the type of spin operators.
We then calculate renormalized spin operators on $d_{e_3^{(i)}} \in d_{\rm max} $ by using Eq.~\eqref{eq:block spin operator}.
Applying the renormalization procedure of Eq.~\eqref{eq:block spin operator} recursively enables to compose further block-spin operators on the edges in $\bm{l}_{\mathfrak{d} - 1}$
with $\{ v_{i} \mid e^{(i)}_3 \in \bm{l}_{\mathfrak{d}} \}$
from $\mathfrak{d} = d_{\rm max}$ to $\mathfrak{d} = 0$.

Using the block-spin operators, we can construct the block Hamiltonian associated with $v_i$ of, for example, the XXZ model of Eq.~\eqref{eq:xxz_hamiltonian}:
\begin{align}\label{eq:block Hamiltonian}
&\tilde{H}_{e^{(i)}_1 e^{(i)}_2} \ = 
\tilde{H}_{e^{(i)}_1} + \tilde{H}_{e^{(i)}_2} + \tilde{H}^{\rm int}_{e^{(i)}_1 e^{(i)}_2} \nonumber~, \\
&\tilde{H}^{\rm int}_{e^{(i)}_1 e^{(i)}_2} = \sum_{r \in {\bm r}^{(i)}_1} \sum_{r' \in {\bm r}^{(i)}_2} \frac{J_{rr'}}{2} \left( \tilde{S}^+_{e^{(i)}_1,r} \tilde{S}^-_{e^{(i)}_2,r'} + \tilde{S}^-_{e^{(i)}_1,r} \tilde{S}^+_{e^{(i)}_2,r'} + \Delta^z_{rr'} \tilde{S}^z_{e^{(i)}_1,r} \tilde{S}^z_{e^{(i)}_2,r'} \right)~.
\end{align}
Then, we get
\begin{equation}\label{eq:projection of Hamiltonian}
\left[\tilde{H}_{e^{(i)}_3}\right]^{i_3}_{i'_3} = \sum_{i_1i_2i'_1i'_2} \Big[v_i\Big]^{i_3}_{i_1i_2} \left[\tilde{H}_{e^{(i)}_1 e^{(i)}_2}\right]_{i_1i_2i'_1i'_2} \Big[v^*_i\Big]^{i'_3}_{i'_1i'_2}~,
\end{equation}
that describes a projection of $\tilde{H}_{e^{(i)}_1 e^{(i)}_2}$ onto a reduced subspace on bond $e^{(i)}_3$ defined by $v_i$.
The superblock Hamiltonian corresponding to two adjacent tensors $\{ v_p, v_q \mid \{p, q\} = \bm{l}_0\}$ surrounding the origin bond $o_{\rm c}$ can be described as:
\begin{align}
\label{eq:superblock Hamiltonian}
\tilde{H}_{e^{(p)}_1 e^{(p)}_2 e^{(q)}_1 e^{(q)}_2} \ &= 
\sum_{k\in\{1,2\}} \sum_{r\in \{p,q\}} \tilde{H}_{e^{(r)}_k} + 
\sum_{r\in \{p,q\}} \tilde{H}^{\rm int}_{e^{(r)}_1 e^{(r)}_{2}}
+ \sum_{k,k'\in\{1,2\}} \tilde{H}^{\rm int}_{e^{(p)}_k e^{(q)}_{k'}} ~.
\end{align}

\subsubsection{Initializing TTN tensors}\label{subsec:initializing tensors}
Given a TTN structure identified by $\bm{E}$, TTNOpt initializes isometric tensors $\bm{v}$ using the real-space renormalization group (RSRG)~\cite{wilsonRenormalizationGroupCritical1975,hikiharaNumericalRenormalizationgroupStudy1999}.
Namely, TTNOpt decides isometries following a recursive sequence from physical bonds to the canonical center of the initial TTN, $o_{\rm c}$, in the same order of composing renormalized spin operators by referring to $\bm {L}$ as introduced in the previous section.

In order to initialize the isometry $v_i$, TTNOpt requires the full diagonalization of $\tilde{H}_{e^{(i)}_1e^{(i)}_2}$ in Eq.~\eqref{eq:block Hamiltonian}.
The corresponding eigenvectors $\bm{u}$ are then collected in ascending order of their eigenvalues, and we obtain the element of isometry by
\begin{equation}
\big[v_i\big]^{i_3}_{i_1i_2} := \big[\bm u\big]^{i_3}_{i_1i_2}~,
\end{equation}
where $ \big[\bm{u}\big]^{i_3}_{i_1i_2} $ is the $ i_3 $-th eigenvector reshaped into a two-dimensional tensor indexed by $ i_1 $ and $ i_2 $.
The maximum bond dimension here is $ \chi_{\rm{init}}$, which is determined by \textbf{numerics.initial\_bond\_dimension}.
Consequently, the computational cost of full diagonalization of the block Hamiltonian is up to $O(\chi_{\rm{init}}^6)$.

It is worth noting that TTNOpt selects up to $\chi_{\rm{init}}$ eigenvectors, accounting for degeneracies arising from symmetries of the system~\cite{whiteDensityMatrixFormulation1992,whiteDensitymatrixAlgorithmsQuantum1993}.
In the case of degenerate eigenvectors, they are either fully retained or discarded to maintain symmetry.
To detect degeneracies, TTNOpt calculates the L1 norm of energy differences  
\begin{equation}  
\Delta_{k} = \big| {\rm{e}}^{k+1}_{e^{(i)}_3} - {\rm{e}}^{k}_{e^{(i)}_3} \big| ,  
\end{equation}  
where $ {\rm{e}}^{k}_{e^{(i)}_3} $ is the $ k $-th eigenvalue of the block Hamiltonian  
$ \tilde{H}_{e^{(i)}_1e^{(i)}_2} $ for $ k $ ranging from $ \chi_{\rm{init}} - 1 $ to $ 0 $.
If $ \Delta_k < \delta_{\rm E} $, where $ \delta_{\rm E} $ is set by  
\textbf{numerics.energy\_degeneracy\_threshold},  
TTNOpt discards the $ (k+1) $-th eigenvector and updates $ k := k - 1 $.
The iteration continues until $\Delta_k \geq \delta_{\rm E}$ and then the practical bond dimension of $i_3$ is determined as $\chi_{e^{(i)}_3} = k+1\in [1,\chi_{\rm init}]$.

Once the RSRG flow is complete, in order to determine singular values tensor $\mathfrak{D}$, TTNOpt derives the renormalized wave function $\tilde{\Psi}_{e^{(p)}_1e^{(p)}_2e^{(q)}_1e^{(q)}_2}$ and bond energy ${\mathfrak E}_{o_{{\rm c}}}$ from the diagonalization of the superblock Hamiltonian $\tilde{H}_{e^{(p)}_1e^{(p)}_2e^{(q)}_1e^{(q)}_2}$ of Eq.~\eqref{eq:superblock Hamiltonian} by using the Lanczos method.
TTNOpt then performs the SVD for $\tilde{\Psi}$ as follows:
\begin{equation}\label{eq:svd}
\Big[\tilde{\Psi}\Big]_{p_1p_2q_1q_2} = \sum_{c}  U^{c}_{p_1p_2} D_{c} V^{c}_{q_1q_2}~,
\end{equation}
where $c$ is up to $\chi_{\rm{init}}^2$.
TTNOpt takes $\chi' \leq \chi_{\rm{init}}$ singular values considering degeneracies of singular values in the same manner as energy degeneracies while, in this case, TTNOpt compares the relative variation $\Delta^{\rm rel}_k = \frac{D_{k+1} - D_{k}}{D_{k}}$, with $k \in [0, \chi_{\rm{init}}-1]$, to $\delta_{\mathcal{S}}$ defined by {\bf numerics.entanglement\_degeneracy\_threshold}.
After the bond dimension $\chi_{o_{\rm c}}$ is decided, TTNOpt replaces two isometries $\{ v_p, v_q \mid \{p, q\} = \bm{l}_0\}$ such that
\begin{align}\label{eq:update canonical center}
    \Big[v_p\Big]^{p_3}_{p_1p_2} &:=  \Big[U\Big]^{p_3}_{p_1p_2} \nonumber \\
    \Big[v_q\Big]^{q_3}_{q_1q_2} &:=  \Big[V\Big]^{q_3}_{q_1q_2}~,
\end{align}
with $p_3,q_3 \in [0, \chi_{o_{\rm c}}-1]$, and we also obtain the normalized singular values $\mathfrak{D}$ with rank $\chi_{o_{\rm c}}$.

In the Lanczos method, $\tilde{H}$ must be applied to a vector $\tilde{\Phi}$ in the truncated Hilbert space as many times as the dimension of the Krylov subspace.
The TTNOpt package calculates $\tilde{H}\tilde{\Phi}$ by applying each term of Eq.~\eqref{eq:superblock Hamiltonian} individually to reduce the computational cost.
Since each operator has a $\chi\times\chi$ size, the computational cost of $\tilde{H}\tilde{\Phi}$ scales as $O(C\chi^5)$, where $C$ is an integer that depends on the number of terms.
To further reduce computational cost, we adjust the order of summation of spin operators.
For example, when taking $\sum_{r\in {\bm{r}}, r'\in {\bm{r}}'}J_{rr'}\tilde{S}^{+}_{e,r}\tilde{S}^{-}_{e',r'}$ with $g' > g$ where $g = |{\bm{r}}|, g' = |{\bm{r'}}|$, we first apply $\sum_{r'\in {\bm{r}}'}J_{rr'}\tilde{S}^{-}_{e',r'}$ to $\tilde{\Phi}$ and subsequently apply $\tilde{S}^{+}_{r}$ for each $r\in \bm{r}$.
Additionally, TTNOpt utilizes $\Big[\tilde \Phi\Big]_{p_1p_2q_1q_2} = \sum\limits_c \big[v_p\big]^{c}_{p_1p_2}\big[v_q\big]^{c}_{q_1q_2}/\sqrt{\sum\limits_{p_1p_2q_1q_2cc'}\big[v_p\big]^{c}_{p_1p_2}\big[v_q\big]^{c}_{q_1q_2}\big[v^*_p\big]^{c'}_{p_1p_2}\big[v^*_q\big]^{c'}_{q_1q_2}}$, where $v_p$ and $v_q$ are decided by the RSRG previously applied, as the initial state for the Lanczos method.

\subsubsection{Main procedure}
\label{subsec:main procedure}
We first show the high-level procedure of the ground state search method of TTNOpt in Algorithm~\ref{algorithm:main}.
Given a TTN state with the mixed canonical form described in Sec.~\ref{subsec:representation of TTN} with above $\bm v$, $\bm E$, $e_{\rm c}$, and $\mathfrak{D}$, TTNOpt updates TTN states based on the two-tensor update method within a sweep procedure as shown in Algorithm~\ref{algorithm:run_sweep}.
Although the path of the sweep is not unique, it has to pass through all tensors in TTN states at least once during a sweep, even if TTN structures are not fixed.
In TTNOpt, we implemented one variety of sweep procedures proposed in Ref.~\cite{hikiharaAutomaticStructuralOptimization2023}.

To illustrate Algorithm~\ref{algorithm:run_sweep}, let us introduce the set of edge labels $\bm e = \{ 0, 1, \cdots, 2N_{\rm t} \}$.
We assign a flag $f_e \in \{0, 1\}$  to each bond $e\in \bm e$ to track the path and completion of the sweep.
In the algorithm, flags for the bonds connecting to bare sites are initialized $1$, i.e., $f_e = 1$ with $e \in [ 0,  N-1 ]$.
We also use the bond distances $d_e$ from the origin bond $o_{\rm c}$.
The distance $d_e$ is decided based on the Breadth-First Search algorithm described in Algorithm~\ref{algorithm:set_distance}.
Furthermore, we define two terminologies: the set of block spin operators $\tilde{\bm S}$ for all bonds and the set of block Hamiltonians $\tilde{\bm{H}}$ for all isometries.
Both of these have been initially constructed during the RSRG procedure.
In the TTNOpt package, $\tilde{\bm S}$ and $\tilde{\bm{H}}$ are implemented as dictionaries.  
Here, $\tilde{\bm{H}}$ stores the block Hamiltonians $\tilde{H}_{e^{(i)}_1e^{(i)}_2}$ with a key $e^{(i)}_3$ for $i \in [0, N_{\rm t} - 1]$ according to the isometry $v_i$.  
Meanwhile, $\tilde{\bm S}$ is a nested dictionary whose elements are indexed by an edge label key $e \in \bm e$.  
Each $\tilde{\bm S}_e$ stores a dictionary of block spin operators $\tilde{S}^{(\cdot)}_{e, r}$ with $(\cdot) \in \{z, +\}$, where the key $r \in {\bm{r}}_e$ represents a set of physical spin locations associated with $e$.

In Algorithm~\ref{algorithm:run_sweep}, the sweep procedure continues until all flags of bonds incoming to the renormalized region of the current canonical center $e_{\rm c}$ are equal to $1$.  
It is worth mentioning that in our algorithm, this situation always happens when $e_{\rm c}$ returns to the origin bond $o_{\rm c}$ and all flags, except for $f_{o_{\rm c}}$, are equal to $1$.
Algorithm~\ref{algorithm:candidate_edge_indices} is used to detect the edges incoming to the renormalized region whose flags are $0$.  
From these edges, we choose the edge $e'_c$, which will be the next canonical center, by using Algorithm~\ref{algorithm:local_two_tensor}.

\begin{figure}[htbp]
  \centering
  \includegraphics[clip, width=3.3in]{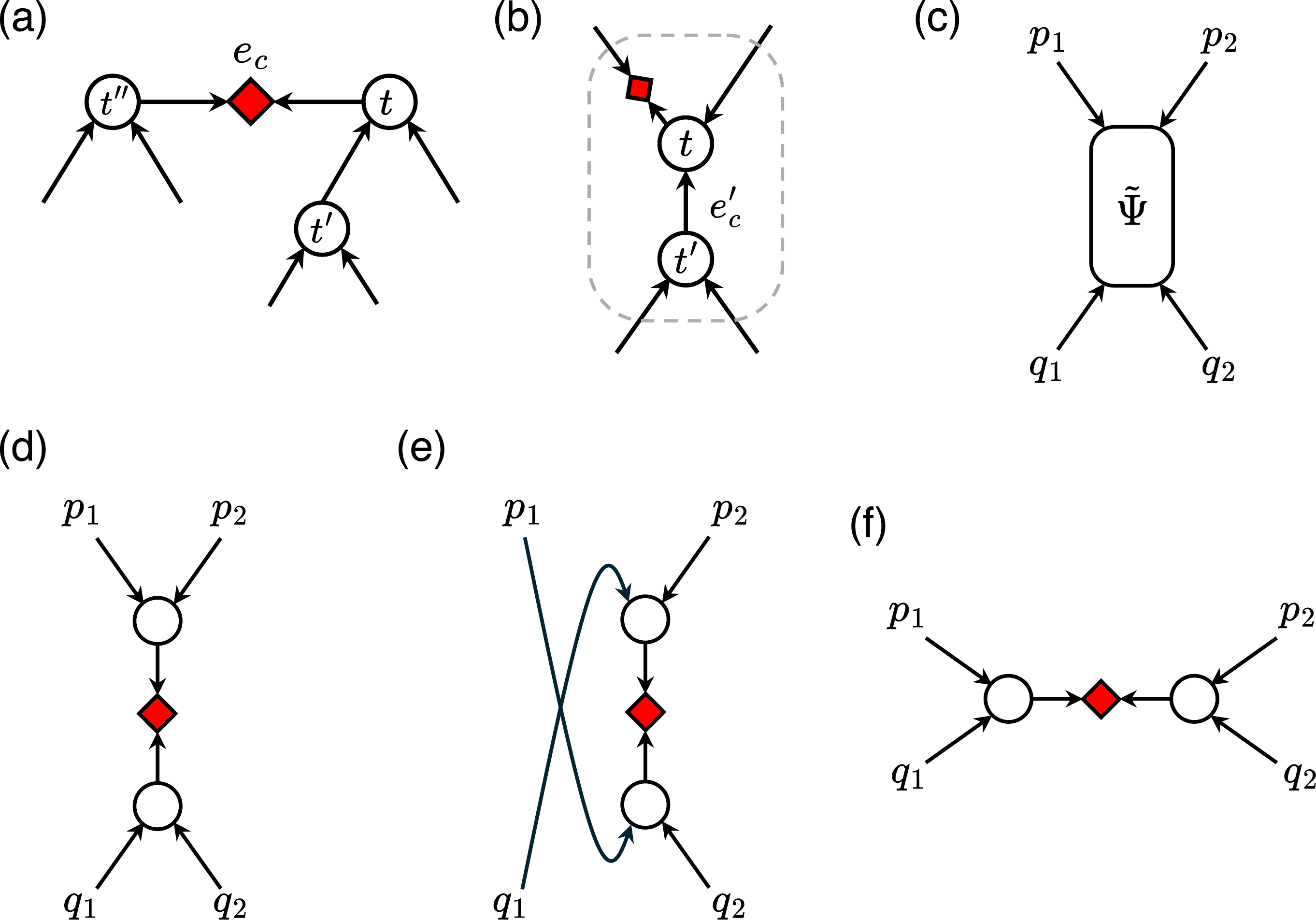}
  \caption{Schematic diagrams of procedures of the two-site tensor update: (a) the step of selecting the next canonical center denoted as $e'_c$ decided by using Algorithm~\ref{algorithm:local_two_tensor},
  (b) the step of contraction within the next canonical region detailed in lines 12 and 13 of Algorithm~\ref{algorithm:run_sweep}, where a gray rounded rectangle represents the region of contracted tensors, 
  (c) renormalized wave function $\tilde{\Psi}$ resulting from the step (b) that is updated by using the Lanczos method in the ground state search,
  and (d),(e), and (f) three possible candidates of SVD of $\tilde{\Psi}$ corresponding to Eq.~\eqref{eq:entanglements}.}
  \label{fig:sweep step}
\end{figure}

Let us assume that the bond $e'_{\rm c}$ connects $v_t$ and $v_{t'}$, as shown in Fig.~\ref{fig:sweep step}$(a)$, and $v_{t''}$ was updated in the previous step.
The flagging process in Algorithm~\ref{algorithm:run_sweep} ensures that $f_{e_{\rm c}} = 1$ only if all bonds in the subtree rooted at the parent tensor $v_{t''}$ have a flag of $1$.
To update tensors $v_t$ and $v_{t'}$, we have to obtain the renormalized wave function $\tilde{\Psi}_{e^{(t)}_1e^{(t)}_2e^{(t')}_1e^{(t')}_2}$ by using the Lanczos method.
We note that TTNOpt contracts $v_t$, $v'_t$, and $\mathfrak{D}$ at $e_c$ as described in Figs.~\ref{fig:sweep step}(a) and (b), and this contracted tensor is used in the Lanczos method as an initial renormalized wave function.
The Lanczos method requires to compose the superblock Hamiltonian $\tilde{H}_{e^{(t)}_1e^{(t)}_2e^{(t')}_1e^{(t')}_2}$ according to the new canonical region specified by $e_{\rm c}'$.
Recall that the effective Hamiltonians and the block spin operators for all bonds except those at $e_{\rm c}$ are retained in $\tilde{\bm{H}}$ and $\tilde{\bm{S}}$, respectively.
This means that only $\tilde{\bm S}_{e_{\rm c}}$ and $\tilde{\bm{H}}_{e_{\rm c}}$ are refreshed by applying $v_{t''}$ in Eqs.~\eqref{eq:block spin operator} and \eqref{eq:block Hamiltonian}.

\begin{figure}[tbp]
\begin{algorithm}[H]
  \caption{Main procedure of ground state search.}
  \label{algorithm:main}
   \begin{algorithmic}[1]
    \State {\textbf{Input:} $H$: the definition of the Hamiltonian, $\bm{E}$: the list of three-integer tuples representing edge labels, $\bm{e}$: the list of integers of edge labels, $o_{\rm c}$: the integer referring to the edge label of the origin, $\mathfrak{m}$: the number of stages of calculations, $\chi_{\rm{init}}$: the maximum bond dimension for initializing tensor, $\bm{\chi}$: the maximum bond dimensions for each stage of calculations, $\bm{n}_{\max}$: the maximum number of sweeps for each stage of calculations, $\epsilon_{\rm E}$: the threshold for the energies, $\epsilon_{\mathcal{S}}$: the shoreshold for the EEs, $l$: the number of consecutive times TTNOpt detects the TTN state as converged before terminating the optimization. This value is set to $2$ by default. }
    \Function{main}{$H, \bm{E}, \bm{e}, o_{\rm c}, \mathfrak{m}, \chi_{\rm{init}}, \bm{\chi}, \bm{n}_{\max}, \epsilon_{\rm E}, \epsilon_{\mathcal{S}}, l$}
        \State $\bm{v}, \mathfrak{D}, \bm{\tilde{S}}, \bm{\tilde{H}} :=$ {\sc{initialize\_ttn}} $(H, \bm{E}, o_{\rm c}, \chi_{\rm{init}})$
      
        \Comment{See Sec.~\ref{subsec:initializing tensors}}
        \For{$m = 1$ to $\mathfrak{m}$}
            \State $c:=0$
            \For{$n = 1$ to $n_{\max,m}$}
                \State $\bm{E}, \bm{v}, \mathfrak{D}, \bm{\tilde{S}}, \bm{\tilde{H}}, \mathfrak{E}, \mathfrak{S} :=${\sc{sweep}}$(\bm E, \bm{e}, o_{\rm c},\bm{v}, \mathfrak{D}, \tilde{\bm{S}}, \tilde{\bm{H}},\chi_m)$
                \Comment{
                $\mathfrak{E}, \mathfrak{S}$ are sets of bond energies and EEs.
                }
                
                \If{$n>1$}
                \If{${\bm{E}_b} = \bm{E}_b'$ for $b \in [0, N_{\rm t}-1]$}
                \If{$|1 - \frac{\mathfrak{E}_b}{\mathfrak{E}'_b}| < \epsilon_{\rm E}$ for $b \in [N, 2N_{\rm t}]$}
                \If{$|\mathfrak{S}_b - \mathfrak{S}'_b| < \epsilon_{\mathcal{S}}$ for $b\in[0,2N_{\rm t}]$}
                \State $c := c + 1$
                    \If{$c > l$}
                        \State Break
                    \EndIf
                \EndIf    
                \EndIf
                \Else
                    \State $c := 0$
                \EndIf
                \EndIf
                \State $\bm{E}' := \bm{E}$
                \State $\mathfrak{E}':=\mathfrak{E}$
                \State $\mathfrak{S}':=\mathfrak{S}$
            \EndFor
        \EndFor
    \EndFunction
   \end{algorithmic}
\end{algorithm}
\end{figure}

\begin{figure}[tbp]
\begin{algorithm}[H]
    \caption{Sweep procedure for the ground state search}
    \label{algorithm:run_sweep}
    \begin{algorithmic}[1]
        \State {\textbf{In/Output:} $\bm{E}, \bm{e}$: the variables introduced in Algorithm.\ref{algorithm:main}, $\bm{v}$: the list of isometric tensors, ${\mathfrak{D}}:$ the normalized vector containing up to $\chi$ singular values, $\tilde{\bm S}:$ the set of block spin operators for all edges, and $\tilde{\bm{H}}:$ the set of block Hamiltonian for all isometries.}
        \State {\textbf{Input:} $o_{\rm c}$: the variables introduced in Algorithm.\ref{algorithm:main}, and $\chi$: the maximum bond dimension of TTN.
        \State {\textbf{Output:} $\mathfrak{E}$: the set of bond energies obtained by the Lanczos method, and $\mathfrak{S}$: the set of EEs obtained from Eq.~\eqref{eq:entanglements}}.}
        \Function{sweep}{$\bm E, \bm{e}, o_{\rm c}, \bm{v}, \mathfrak{D}, \tilde{\bm{S}}, \tilde{\bm{H}}, \chi$}
        \State $e_{\rm c} := o_{\rm c}$
        \State $\bm{f} := \{0 \mid f_e, {\rm where }~e \in \bm{e} \}$
        \State $\bm{d} := $ {\sc{set\_distance}}$(\bm{E}, \bm e, e_{\rm c})$
        \State $\mathfrak{E} := \{ \}$
        \Comment{Initialize the set of energies $\mathfrak{E}$.}
        \State $\mathfrak{S} := \{ \}$
        \Comment{Initialize the set of EEs $\mathfrak{S}$.}
        \While{ {\sc candidate\_edge\_indices}$(\bm{E}, e_{\rm c}, \bm f) \neq \{\}$ }
        \State $e'_{\rm c}, t, t', t'' :=$ {\sc{local\_two\_tensor}}$(\bm{E}, e_{\rm c}, \bm{f}, \bm{d})$

        \Comment{ See Fig.~\ref{fig:sweep step}(a).}
        \If {$\left\{f_e = 1 \mid e \in \{ e^{(t'')}_1, e^{(t'')}_2\}\right\}$}
        
        \Comment{Recall that $E_{t''} = (e^{(t'')}_1, e^{(t'')}_2, e^{(t'')}_3)$.}
            \If {$e_c \neq o_{\rm c}$}
            \State $f_{e_{\rm c}} := 1$ \Comment{At this point, it satisfies $e_{\rm c} = e^{(t'')}_{3}$}
            \EndIf
        \EndIf
        \Statex \Comment If $f_{e_{\rm c}}$ becomes $1$, it calculates expectation values according to $v_{t''}$ with $\tilde{\Psi}$ obtained at the previous step.
        \State Update $\tilde{\bm S}_{e_{\rm c}}$ by Eq.~\eqref{eq:block spin operator} with $v_{t''}$
        \State Update $\tilde{\bm {H}}_{e_{\rm c}}$ by Eq.~\eqref{eq:block Hamiltonian} with $v_{t''}$
        \State $\tilde\Psi := v_{t} \circ {\mathfrak{D}} \circ v_{t'}$        
        \Statex \Comment $\circ$ denotes the contraction of tensors according to the same indices based on $\bm{E}_t$ and $\bm{E}_{t'}$. See Figs.~\ref{fig:sweep step} (b) and (c).
        \State $\tilde\Psi, {\rm E} := ${\sc{lanczos}}$(\tilde\Psi, \tilde{\bm S}, \tilde{\bm{H}})$
        \State $\mathfrak{E}_{e'_{\rm c}} := \rm{E}$
        
        \State $v_t, {\mathfrak{D}}, v_{t'}, \mathcal{S} :=$ {\sc{decompose\_tensor}} $(\tilde{\Psi}, \chi)$
        \State $\mathfrak{S}_{e'_{\rm c}} := \mathcal{S}$
        \Comment $\mathcal{S}$ is calculated by Eq.~\eqref{eq:entanglements}.
        \For{$r \in \{ e^{(t)}_1,e^{(t)}_2,e^{(t')}_1,e^{(t')}_2\}$}
            \If{$r \in [0, N-1]$}
                \State $\mathcal{S}_{r} := $ {\sc{site\_ee}} $(\tilde{\Psi}, r)$
                \Comment{See Fig.~\ref{fig: site ee}.}
                \State $\mathfrak{S}_{r} := \mathcal{S}_{r}$
            \EndIf            
        \EndFor
        \State $e_{\rm c} := e'_{\rm c}$
        \State Update $E_t, E_{t'}$
        \State $\bm{d} := $ {\sc{set\_distance}}$(\bm{E}, \bm e, o_{\rm c})$
        \EndWhile

        \Comment{ Expectation values are calculated using $v_{t}$ and $v_{t'}$ which construct $\tilde{\Psi}$ obtained at the last step.}

        \Return $\bm{E}, \bm{v},\bm{e}, \mathfrak{D}, \bm{\tilde{S}}, \bm{\tilde{H}}, \mathfrak{E}, \mathfrak{S}$
        \EndFunction
    \end{algorithmic}
\end{algorithm}
\end{figure}

\begin{figure}[htbp]
\centering
\includegraphics[clip, width=3.3in]{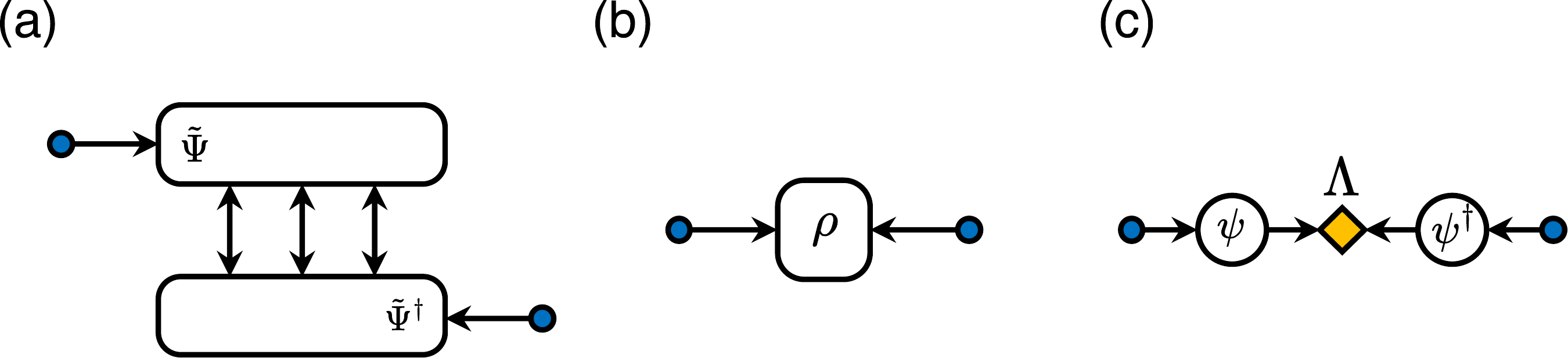}
\caption{Schematic diagrams of the process of the {\sc{site\_ee}} function.
(a) and (b) represent the calculation of $\Big[\rho\Big]_{p_1p'_1} = \sum\limits_{p_2q_1q_2}\Big[\tilde \Psi \Big]_{p_1p_2q_1q_2}\Big[\tilde{\Psi}^* \Big]_{p'_1p_2q_1q_2}$, where $p_1$ here is the index for a bare site $s_r$ with $r \in [0, N-1]$.
(c) describes the diagonalization for $\rho$ to obtain the EE on the bond $p_1$, i.e., $\Big[\rho\Big]_{p_1p'_1} = \sum_c\Big[{\psi}\Big]_{p_1c}\Big[\Lambda \Big]_{c}\Big[\psi^* \Big]_{cp'_1}$ and $\mathcal{S} = -\sum_{c}\Lambda_{c}\ln{\Lambda_{c}}$.
}
\label{fig: site ee}
\end{figure}

After obtaining $\Big[\tilde{\Psi}_{e^{(t)}_1 e^{(t)}_2e^{(t')}_1e^{(t')}_2}\Big]_{p_1p_2q_1q_2}$, where $\{p_1, p_2, q_1, q_2\}$ are indices corresponding to edge labels $\{e^{(t)}_1, e^{(t)}_2, e^{(t')}_1, e^{(t')}_2 \}$, the {\sc{decompose\_tensor}} function is performed to update tensors and local structure.
For the reconnection of $\tilde\Psi_{e^{(t)}_1 e^{(t)}_2e^{(t')}_1e^{(t')}_2}$, there exist three possible index orders: $(p_1p_2 \mid q_1q_2)$, $(p_1q_2 \mid p_2q_1)$, and $(p_1q_1 \mid p_2q_2)$ as shown in Figs.~\ref{fig:sweep step}(d), (e), and (f).
In {\sc{decompose\_tensor}} function, the EEs for all three configuration are computed by performing a full SVD, which has a computational cost of $O(\chi^6)$.
The EEs are given by
\begin{align}\label{eq:entanglements}
    \mathcal{S}^{(p_1p_2\mid q_1q_2)} &= - \sum_{c} (D_c)^2 \ln{(D_c)^2}, \nonumber \\
    \mathcal{S}^{(p_1q_2 \mid p_2q_1)} &= - \sum_{c} (D'_c)^2 \ln{(D'_c)^2}, \nonumber \\
    \mathcal{S}^{(p_1q_1 \mid p_2q_2)} &= - \sum_{c} (D''_c)^2 \ln{(D''_c)^2}, 
\end{align}
where
\begin{align}
\Big[\tilde\Psi_{{e^{(t)}_1 e^{(t)}_2e^{(t')}_1e^{(t')}_2}}\Big]_{p_1p_2q_1q_2} 
        &= \sum_{c} \Big[U\Big]^c_{p_1p_2} \Big[D\Big]_c \Big[V\Big]^c_{q_1q_2}, \nonumber \\
    &= \sum_{c} \Big[U'\Big]^c_{p_1q_2}\Big[D'\Big]_c \Big[V'\Big]^c_{p_2q_1}, \nonumber \\
    &= \sum_{c} \Big[U''\Big]^c_{p_1q_1} \Big[D''\Big]_c \Big[V''\Big]^c_{p_2q_2},
\end{align}
respectively and $c$ is up to $\chi^2$.
TTNOpt then selects the structure with the smallest EE, if users set {\bf{opt\_structure.type}} to $1$.
However, if the minimum EE $\mathcal{S}_{\rm min}$ and $\mathcal{S}^{(p_1p_2|q_1q_2)}$ of the original structure [Figs.~\ref{fig:sweep step}(d)], satisfy the condition  
$
|\mathcal{S}^{(p_1p_2 \mid q_1q_2)} - \mathcal{S}_{\rm min}| < \epsilon_{\mathcal{S}}
$,
TTNOpt retains the original connection to avoid insignificant variations in the TTN structure.
Here, $\epsilon_{\mathcal{S}}$ is defined by {\bf numerics.entanglement\_convergence\_threshold}.
Once the optimal structure is determined, TTNOpt truncates any singular values exceeding the rank $\chi$ set by {\bf numerics.max\_bond\_dimension}, while accounting for degeneracies in the singular values with $\delta_{\mathcal{S}}$.

Since the reconnection procedure employed in TTNOpt is local, the solution may be trapped in local minima, especially in complex systems such as disordered ones~\cite{hikiharaImprovingAccuracyTreetensor2025}.
To overcome this problem, TTNOpt has a function to select a structure based on relative probabilities.
The heat-bath method works by evaluating the EEs for the three possible reconnections and sampling the one with a distribution given by the following expression
\begin{align}\label{eq:stochastic selection}
P^{(p_1p_2 \mid q_1q_2)} &\propto \exp \left[-\mathcal{S}^{(p_1p_2 \mid q_1q_2)} / T \right]~, \nonumber \\
P^{(p_1q_2 \mid p_2q_1)} &\propto \exp \left[- \mathcal{S}^{(p_1q_2 \mid p_2q_1)} / T\right]~, \nonumber \\
P^{(p_1q_1 \mid p_2q_2)} &\propto \exp \left[-\mathcal{S}^{(p_1q_1 \mid p_2q_2)} / T\right]~,
\end{align}
with
\begin{equation}\label{eq:manipulate temperature}
    T=2^{-n/n_{\tau}}T_0~,
\end{equation}
where an initial temperature $T_0$ and a decay factor $n_{\tau}$ are set by {\bf numerics.opt\_structure.temperature} and {\bf numerics.opt\_structure.tau}, respectively, and $n \in [0, n_{\rm{max}}-1]$ represents the sweep number with $n_{\rm{max}}$ assigned by {\bf numerics.max\_sweep\_nums}.
To achieve the convergence of the TTN structure during the sweep process, TTNOpt exponentially decreases $T$ to $0$ (see Ref.~\cite{hikiharaImprovingAccuracyTreetensor2025}).

Furthermore, TTNOpt can select the structure with the minimum truncation error if {\bf{opt\_structure.type}} is set to $2$.
Here, the truncation error is defined as
\begin{equation}\label{eq:truncated singular values}
    \Delta = 1 - \sum^{\chi}_{c=1} (D_c)^2~.
\end{equation}
for $(p_1p_2 \mid q_1q_2)$ and similarly for the other decompositions.

As shown in Algorithm~\ref{algorithm:main}, the {\sc{run\_sweep}} is repeated until the number of sweeps reaches $n_{\max}$ set by ${\bf{numerics.max\_num\_sweeps}}$ or the structure described by $\bm E$, variational energies, and EEs have been converged.
TTNOpt saves the variational energy obtained by the Lanczos diagonalization performed at auxiliary bonds in the set $\mathfrak{E}$, as well as the EEs on all bonds, including physical ones, in the set $\mathfrak{S}$. 
These values are always overwritten for the same bonds in the sweep.
To check for convergence, TTNOpt computes the difference in variatonal energies and EEs for all considered bonds with those from the previous sweep, and judges the convergence with $\epsilon_{\rm E}$ and $\epsilon_{\mathcal{S}}$ specified by {\bf{numerics.energy\_convergence\_threshold}} and {\bf{numerics.entanglement\_convergence\_threshold}}, respectively.

\begin{figure}[t]
\begin{algorithm}[H]
  \caption{Set distances from the edge $e$ using the Breadth-First Search.}
  \label{algorithm:set_distance}
   \begin{algorithmic}[1]
    \Require{$\bm{E}, \bm{e}, o_{\rm c}$: the variables introduced in Algorithm.\ref{algorithm:main}.}
    \Ensure{$\bm{d}$: the list of distances from edge $o_{\rm c}$}
    
    \Function{set\_distance}{$\bm{E}, \bm{e}, o_{\rm c}$}
      \For {$e \in \bm{e}$}
        \State $\bm{A}_e:=\bigcup_{\mathcal{E} \in \boldsymbol{E}}\{e' \mid e \in \mathcal{E}, e' \in \mathcal{E}, e' \neq e\}$
      \EndFor
      \State $\bm d := \{ 0 \mid d_e, {\rm{where}}~e \in \bm{e}\}$ \Comment{Initialize entries of distance list as $0$}
      \State Initialize an empty queue $Q$
      \State Enqueue $o_{\rm c}$ into $Q$
      \While {$Q$ is not empty}
      \State Dequeue $e'$ from $Q$
      \State $d' := d_{e'}$
      \For{each neighbor edge $u$ of $e'$ in $\bm A_{e'}$}
      \If{$d_u = 0$}
      \State $d_u := d_{e'} + 1$
      \State Enqueue $u$ into $Q$
      \EndIf
      \EndFor
      \EndWhile
      \State \Return $\bm{d}$
    \EndFunction
   \end{algorithmic}
\end{algorithm}
\end{figure}

\begin{figure}[H]
\begin{algorithm}[H]
  \caption{Detect candidate edges.}
  \label{algorithm:candidate_edge_indices}
   \begin{algorithmic}[1]
    \Require{$\bm{E}$: the valiable introduced in Algorithm~\ref{algorithm:main}, $e_{\rm c}$: the integer of edge label of the current canonical center, and $\bm{f}$: the list of bools of flag at each edge label.}
    \Ensure{$\bm{c}$: the list of edge indices}
    \Function{candidate\_edge\_indices}{$\bm E, e_{\rm c}, \bm f$}
    \State $\bm{c} :=\bigcup_{(\mathcal{E}_1,\mathcal{E}_2,\mathcal{E}_3) \in \boldsymbol{E}}\{e \mid e \in \{ \mathcal{E}_1, \mathcal{E}_2 \}, \mathcal{E}_3 = e_{\rm c}\}$ \Comment{Find edges adjescent to canonical center.}
    \State $\bm{c} := \{i \in \bm{c} \mid f_i = 0\}$ \Comment{Filter by flag $\bm{f}$.}
    \State \Return $\bm{c}$
    \EndFunction
   \end{algorithmic}
\end{algorithm}
\end{figure}

\begin{figure}[t]
\begin{algorithm}[H]
  \caption{Select two local tensors connected by the edge with the largest distance from the initial canonical center.}
  \label{algorithm:local_two_tensor}
   \begin{algorithmic}[1]
\Require{$\bm{E}, e_{\rm c}, \bm{f}$: the variables introduced in Algorithm~\ref{algorithm:candidate_edge_indices}, and $\bm{d}$: the list of integers of distance at each edge}
    \Ensure{$e_{\rm c}'$: the integer of edge label of the next canonical center, $t, t', t''$: the integers of tensor labels.}
    
    \Function{local\_two\_tensor}{$\bm{E}, e_{\rm c}, \bm{f}, \bm{d}$}
      \State $\bm{c} :=$ \Call{candidate\_edge\_indices}{$\bm{E}, e_{\rm c}, \bm{f}$}
      \State $d_{\max} := \max(\{d_e \mid e \in \bm{c}\})$
      \State $\bm{c} := \{e \in \bm{c} \mid d_e = d_{\max}\}$
      \State $e'_{\rm c} := c_1$ \Comment{Select the first edge in $\bm{c}$}
      \For{$i\in [0, N_{\rm t}-1]$}
        \If{$e_{\rm c} \in E_i \land e'_{\rm c} \in E_i$}
            \State $t := i$
        \EndIf
        
        \If{$ e_{\rm c} \notin E_i \land e'_{\rm c} \in E_i$}
            \State $t' := i$
        \EndIf

        \If{$e_{\rm c} \in E_i \land e'_{\rm c} \notin E_i$}
            \State $t'' := i$
        \EndIf 
      \EndFor
      \State \Return $(e'_{\rm c}, t, t', t'')$
    \EndFunction
   \end{algorithmic}
\end{algorithm}
\end{figure}

\subsubsection{Calculating expectation values}\label{subsec: calculating expectation values}
The TTNOpt package computes expectation values for one and two-site spin operators in sweep procedures using block spin operators $\tilde{\bm S}$.
To ensure that the computation covers all bare sites and site pairs without duplication or omission, TTNOpt assumes the structure remains unchanged during the calculation of the expectation values.
Therefore, if users specify the structural optimization conducted, TTNOpt performs an additional sweep to calculate the expectation values after the first update stage, where $m=1$, for the TTN with the optimized structure.
In the stages of $m>1$, the same calculations are carried out in every sweep since the structure is fixed.

TTNOpt calculates expectation values using $\tilde{\Psi}_{e^{(t)}_1e^{(t)}_2e^{(t'')}_1e^{(t'')}_2}$ in Fig.~\ref{fig:sweep step}(a) at each step.
Single-site expectation values are computed when the renormalized wave function $\tilde\Psi$ is directly associated with the corresponding physical site.
Two-site expectation values are evaluated at the step where the edge of the canonical center, $e'_c$, lies on the minimal path connecting the two physical sites.
See Algorithm~\ref{algorithm:main} for the details.

Let us denote for simplicity the renormalized wave function as $\Big[\tilde{\Psi}\Big]_{p_1p_2q_1q_2}$ with $(p, q) = (t'', t)$ and eliminate the subscripts $\{e^{(t)}_1, e^{(t)}_2, e^{(t'')}_1, e^{(t'')}_2\}$.
When $v_{p}$ directly connects with the physical site $r$, expectation values of spin operators for the site are evaluated by the following equation
\begin{equation}\label{eq:one site expectation value}
\expval{s^{\alpha}_r} = 
\left\{
\begin{array}{ll}
\sum\limits_{p_1p_2q_1q_2p'_1} \Big[ \tilde{\Psi}^{*} \Big]_{p_1p_2q_1q_2}\Big[ \tilde{\Psi} \Big]_{p'_1p_2q_1q_2}
\left[s^{(\alpha)}_r\right]_{p_1p'_1} (r = e^{(p)}_1)\\

\sum\limits_{p_1p_2q_1q_2p'_2} \Big[ \tilde{\Psi}^{*} \Big]_{p_1p_2q_1q_2}
\Big[ \tilde{\Psi} \Big]_{p_1p'_2q_1q_2}\left[s^{(\alpha)}_r\right]_{p_2p'_2} (r = e^{(p)}_2)
\end{array}
\right. ~,
\end{equation}
with $\alpha \in \{ x, y, z \}$.
The two-site correlation between sites $r \in {\bm r^{(p)}_1}$ and $r' \in {\bm r^{(p)}_2}$ are obtained by
\begin{equation}\label{eq:two site correlation}
\expval{s_r^{\alpha}s_{r'}^{\beta}} = 
\sum_{p_1p_2q_1q_2p'_1p'_2} 
\Big[ \tilde{\Psi}^{*} \Big]_{p_1p_2q_1q_2}
\Big[ \tilde{\Psi} \Big]_{p'_1p'_2q_1q_2}
\Big[ \tilde{S}^{(\alpha)}_{e^{(p)}_1, r} \Big]_{p_1p'_1}
\Big[ \tilde{S}^{(\beta)}_{e^{(p)}_2, r'} \Big]_{p_2p'_2}
~,
\end{equation}
with $(\alpha, \beta) \in \{x,y,z\}^{\otimes2}$.

After completing a sweep, TTNOpt calculates expectation values concerning the origin bond $o_{\rm c}$ with the renormalized wave function $\tilde{\Psi}_{e^{(t)}_1e^{(t)}_2e^{(t')}_1e^{(t')}_2}$ obtained at the final step of sweeps as shown in Fig.~\ref{fig:sweep step}(b), (c).
We denote $\Big[\tilde{\Psi}_{e^{(t)}_1e^{(t)}_2e^{(t')}_1e^{(t')}_2} \Big]_{p_1p_2q_1q_2}$ as $\Big[\tilde{\Psi}\Big]_{p_1p_2q_1q_2}$ eliminating the subscripts with $(p, q) \in \{(t, t'), (t', t)\}$.
Single site expectation values are obtained using the same equation as Eq.~\eqref{eq:one site expectation value}, which is calculated only in the case 
that one (or some) of the bonds $\{e_1^{(p)}, e_2^{(p)}, e_1^{(q)}, e_2^{(q)}\}$ is connected directly with a physical site $r$.
Regarding two-site correlations, TTNOpt calculates expectation values for all spin pairs that have not yet been evaluated during the sweeps, using the renormalized wave function associated with the pair $(r, r')$,
where $r \in \{ {\bm r}^{(p)}_{1}, {\bm r}^{(p)}_{2} \}$ and $r' \in \{ {\bm r}^{(q)}_{1}, {\bm r}^{(q)}_{2} \}$.  
For example, in the case of $(r, r') = ({\bm r}^{(p)}_{1}, {\bm r}^{(q)}_{1})$, the corresponding contraction is evaluated as
\begin{align}\label{eq:two site correlation origin}
\expval{s_r^{(\alpha)}s_{r'}^{(\beta)}} = \sum_{\substack{p_1p_2q_1q_2\\p'_1q'_1}}
\Big[\tilde{\Psi}^{*}\Big]_{p_1p_2q_1q_2}
\Big[\tilde{\Psi}\Big]_{p'_1p_2q'_1q_2}
\Big[ \tilde{S}^{(\alpha)}_{e^{(p)}_{1}, r} \Big]_{p_{1}p'_{1}}
\Big[ \tilde{S}^{(\beta)}_{e^{(q)}_{1}, r'} \Big]_{q_{1}q'_{1}}~.
\end{align}

In our algorithm, the expectation values are computed at the step where the distance between the renormalized region in which the wave function $\tilde\Psi$ is obtained and the corresponding physical sites is minimized.
This approach stems from the idea that minimizing the number of renormalization steps for spin operators can reduce the loss of accuracy induced by truncation.
However, it remains an open question whether the current strategy outperforms the alternative approach in which expectation values are evaluated using the fixed TTN wavefunction $\ket{\Psi}$ after the completion of the sweep.

\subsection{Factorizing tensors}\label{subsec:factorizing tensor}
Let us assume that a rank-$N$ tensor $\Psi_{s_0, \cdots, s_{N-1}}$ is given by {\bf target.tensor}, and $\Psi$ is normalized as
\begin{equation}
    \Psi := \frac{\Psi}{\sqrt{\sum\limits_{s_0,\cdots, s_{N-1}} \Big[\Psi\Big]_{s_0, \cdots, s_{N-1}}\Big[\Psi^*\Big]_{s_0, \cdots, s_{N-1}}}}~,
\end{equation}
which is performed by TTNOpt itself before the factorization.
TTNOpt then proceeds to decompose $\Psi$ by using the sequential SVD into the MPN form~\cite{schollwockDensitymatrixRenormalizationGroup2011} with a bond dimension $\chi_{\mathrm{init}}$ specified in {\bf{numerics.initial\_bond\_dimension}} [Fig.~\ref{fig:ft_commands}(c)].
If users set {\bf{numerics.opt\_structure.type}} as $1$ or $2$, TTNOpt runs sweeps to transform the TTN with reconnection of local structures [Fig.~\ref{fig:ft_commands}(d)].
During these sweeps, TTNOpt applies the SVD with $\chi_{\mathrm{init}}$ for the renormalized wave function $\tilde{\Psi}$ composed by the contraction of two tensors, i.e., skipping the Lanczos methods unlike the ground state search in Algorithm~\ref{algorithm:run_sweep}.
This process continues until the TTN structure is fixed and the EEs are converged within the value $\epsilon_{\mathcal{S}}$ specified by {\bf{numerics.entanglement\_convergence\_threshold}}.

If users specify variables of {\bf{numerics.fidelity}}, TTNOpt optimizes the TTN state based on the fidelity with the original tensor $\Psi$.  
TTNOpt can also include the network update in this process, depending on the input file.
To obtain the optimal renormalized wave function $\tilde{\Psi}_{e^{(p)}_1e^{(p)}_2e^{(q)}_1e^{(q)}_2}$, it is necessary to compute the environment $\mathcal{E}_{e^{(p)}_1e^{(p)}_2e^{(q)}_1e^{(q)}_2}$ associated with two tensors $v_{p}$ and $v_{q}$ within the canonical region:
\begin{equation}\label{eq:environment}
\mathcal{E}_{e^{(p)}_1e^{(p)}_2e^{(q)}_1e^{(q)}_2} = \Psi
\prod_{i \in [ 0, N_t-1] / \{ p, q \}} \circ ~v^*_i
~,
\end{equation}
where $\circ$ represents the contraction between tensors with the same indices following $\bm{E}$.
We directly embed the environment as  
\begin{equation}
    \Big[\tilde{\Psi} \Big]_{p_1p_2q_1q_2} := \frac{\Big[\mathcal{E}\Big]_{p_1p_2q_1q_2}}{\sqrt{ \sum\limits_{p_1p_2q_1q_2} \Big[\mathcal{E}\Big]_{p_1p_2q_1q_2} \Big[\mathcal{E}^*\Big]_{p_1p_2q_1q_2}}}~,
\end{equation}
where we omitted the subscripts on tensors, to locally maximize the fidelity with $\Psi$.

The contraction of $\mathcal{E}$ is performed from the physical sites to the canonical center, which ensures that the environment tensor is built step by step while respecting the structure of the network.
TTNOpt iterates the sweeps until the TTN state converges with respect to the network structure, the EE, and the fidelity.
The convergence criteria of EE and fidelity are set by the thresholds $\epsilon_{\mathcal{S}}$ and $\epsilon_F$, specified in {\bf numerics.fidelity.convergence\_entanglement} and {\bf numerics.fidelity.convergence\_threshold}, respectively.

\subsection{Reconstructing TTNs}\label{subsec:reconstruction of TTNs}
If the TTN state is loaded by {\bf{target.tensors}}, users must specify {\bf{numerics.max\_sweep\_num}} and {\bf{numerics.opt\_structure.type}} as either $1$ or $2$.
TTNOpt applies sweeps to the TTN to reconstruct its network by using {\sc{decompose\_tensor}} in the same way as described in the first paragraph of Sec.~\ref{subsec:factorizing tensor} [Fig.~\ref{fig:ft_commands}(d)].
It is emphasized that TTNOpt retains up to $\chi_{\rm{init}}$ singular values during the SVD, where $\chi_{\rm{init}}$ is the maximum bond dimension of the given TTN state.
The calculation terminates when the TTN state has converged concerning both the network structure and the EE, or when the maximum number of sweeps is reached.

\section{Benchmark results}\label{sec:benchmark}
\subsection{Hierarchical chain model}\label{sec:benchmark ground state search}
\begin{figure}[t]
  \centering
  \includegraphics[clip, width=3.3in]{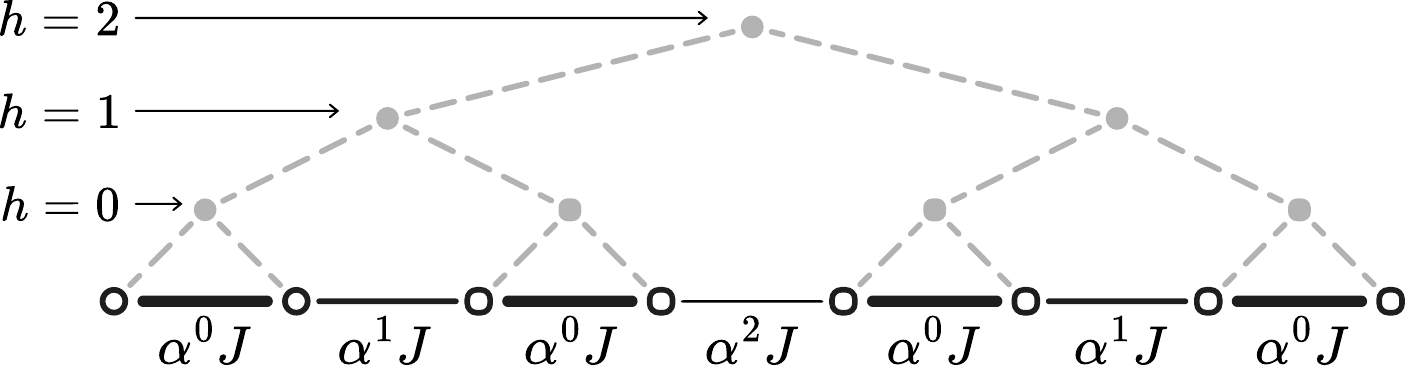}
  \caption{Representation of the hierarchical chain model with $N=8$. The physical sites are ordered from left to right with indices $i=0, \cdots, N-1$. }
  \label{fig:hierarchical_chain}
\end{figure}

To briefly demonstrate the TTNOpt package, we performed the ground state search for the $S=1/2$ hierarchical chain model~\cite{hikiharaAutomaticStructuralOptimization2023} with system size $N=2^{d}$ of an integer $d$, defined as
\begin{equation}\label{eq:Hieralchical chain model}
    H = \sum_{h=0}^{{d}-1} \sum_{i \in I(h)} J \alpha^h {\bm{s}}_i \cdot {\bm{s}}_{i+1}~,
\end{equation}
where $J>0$ is the base coupling constant, and $0 < \alpha \leq 1.0$ is the decay factor for the coupling strength. 
In Eq.~\eqref{eq:Hieralchical chain model}, an integer $h \in [0, d-1]$ represents the height in the perfect binary tree (PBT) structure, and an integer set $I(h)=\left\{i \mid i=2^h(2 k+1)-1,~ {\rm{where}}~k=0, 1, \ldots, 2^{d-h-1}-1\right\}$ specifies pairs of adjacent sites $(i, i+1)$ according to the PBT structure, as illustrated in Fig.~\ref{fig:hierarchical_chain}.
\begin{figure*}[ht]
  \centering
  \includegraphics[clip, width=0.95\textwidth]{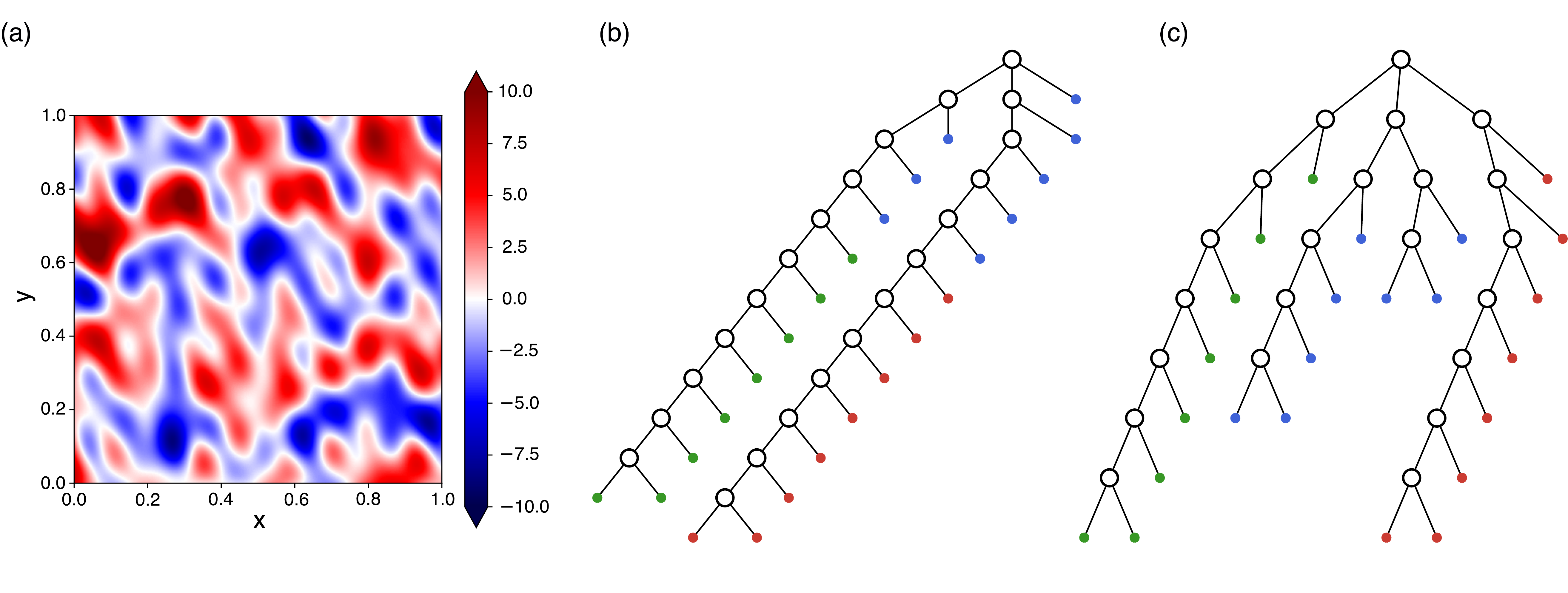}
  \caption{(a) The heatmap of the function Eq.~\eqref{eq:multivariable function} at $x_3=0.5$, where $({\rm x}, {\rm y}):=(x_1, x_2)$. (b) MPN decomposition of the quantics function Eq.~\eqref{eq:multivariable function} where the red, blue, and green circles distinguish variants $x_1$, $x_2$, and $x_3$, respectively. 
  (c) TTN structure obtained by the two-site update algorithm in TTNOpt, which optimizes the structure while maximizing the fidelity with respect to the original functional tensor, starting from the MPN depicted in (b).
  In both (b) and (c), we omit the drawing of the canonical center and arrows for clarity. }
  \label{fig:factorizing tensor}
\end{figure*}

We examined the model with $(J, \alpha)= (1.0, 0.5)$ and $(1.0, 1.0)$, where the system size is $N=256$, i.e., $d=8$.
These settings serve as reasonable litmus tests since $\alpha$ controls the interaction strength between adjacent spins, directly influencing the entanglement structure of the ground states.
It allows for predicting ideal TTN structures: for $\alpha=0.5$, by the perturbative RSRG approach~\cite{PhysRevLett.43.1434}, the optimal TTN structure for the ground state would be the PBT.
For $\alpha=1.0$, the system is the uniform Heisenberg chain under the open boundary conditions, and the optimal TTN structure for the ground state is expected to be an MPN-like one consisting of the dimer units as shown in Ref.~\cite{hikiharaAutomaticStructuralOptimization2023}.

We ran the variational algorithm with the maximum bond dimension $\chi=20$ to find the optimal TTN structures.
Additionally, we set the maximum number of sweeps to $50$ and performed the calculation with $\delta_{\rm E} = 1 \times 10^{-11}$ and $\delta_{\mathcal{S}} = 1 \times 10^{-10}$, respectively.
The remaining parameters used in the calculation can be found in the input file  ``samples/ground\_state\_search/hierarchical''.

We demonstrate that TTNOpt can achieve the same optimal structure for both $\alpha = 0.5$ and $1.0$ as Ref.~\cite{hikiharaAutomaticStructuralOptimization2023}, where the system up to $N=128$ sites was treated.
Furthermore, to show the importance of TTN structures, we present the bond EEs in the optimized TTN and MPN in Table~\ref{Table_I}.
It shows that the optimal TTNs reduced both the maximum and average EEs compared to the case with MPN.
The reduction is significant, especially in the case of $\alpha=0.5$.
On the other hand, in $\alpha=1.0$, it is subtle since the MPN and the dimer MPN are similar structurally.

%
\begin{table}[th]
\caption{Maximum and average of the bond EEs in the optimized TTN and MPN when $N = 256$ with $\alpha = 0.5$ and $1.0$.
The EEs on the physical bonds are excluded from the analysis.}
\label{Table_I}
\begin{center}
\begin{tabular}{l c c}
\hline
\hline
Type &  Maximum & Average \\
\hline
optimized TTN ($\alpha=0.5$) & 0.1110 & 0.0618 \\
MPN ~ ($\alpha=0.5$) & 0.6935 & 0.3730 \\
\hline
optimized TTN ($\alpha=1.0$) & 0.9977 & 0.9065 \\
MPN ~ ($\alpha=1.0$) & 1.0150 & 0.9376 \\
\hline
\hline
\end{tabular}
\end{center}
\end{table}

\begin{figure*}[t]
  \centering
  \includegraphics[clip, width=0.95\textwidth]{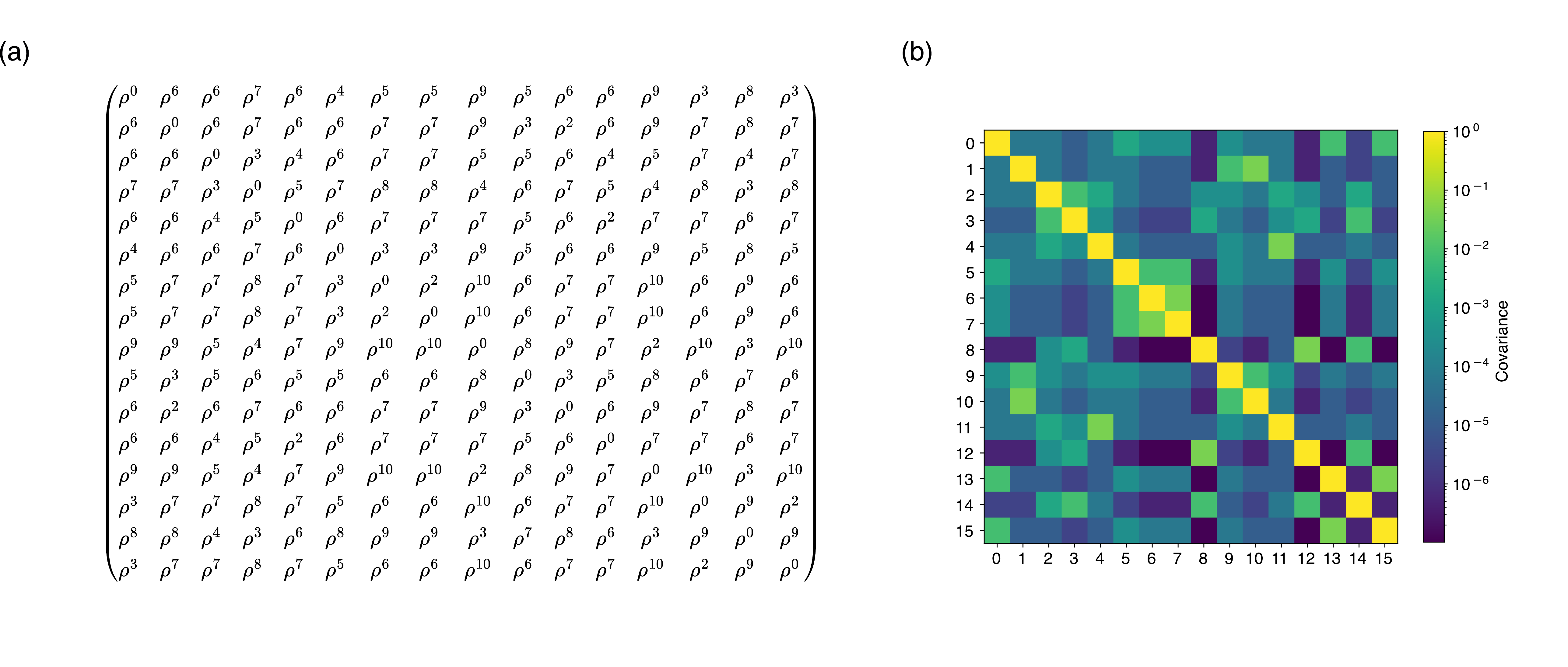}
  \caption{The definition of the covariance matrix $K$ in this experiment with $\rho=0.2$. (b) The heatmap of the matrix elements of $K$ is shown as a reference.}
  \label{fig:cov}
\end{figure*}

\subsection{Multivariable quantics function}\label{sec:benchmark factorize tensor}

Recently, the impact of TTN structures on approximating the general tensor data and compressing quantics tensors has been studied~\cite{tindallCompressingMultivariateFunctions2024}.
Here, we apply TTNOpt to the compression of the three-variable quantics function employed in Ref.~\cite{tindallCompressingMultivariateFunctions2024} that is written as 
\begin{equation}\label{eq:multivariable function}
    f({\bm{x}}) = \sum_{j=1}^n \cos \left(j {\bm{k}}_{j} \cdot {\bm{x}} \right)~,
\end{equation}
where $n = 30$, $\bm{x} = (x_1, x_2, x_3) \in [0, 1)^{\otimes3}$ and $\bm{k}_j = (k_{j,1}, k_{j,2}, k_{j,3})$ with $k^\alpha_j \in \mathcal{N}(0, 1)$ that is the standard normal distribution.
In quantics formulation, the continous variable $x_i \in [0, 1)$  with the $L$-bit precision is described as
\begin{equation}\label{eq:quantics variable}
    x_i = \sum^{L}_{l=1} \frac{x_{i,l}}{2^{l}}~,
\end{equation}
where $x_{i,l} \in \{ 0, 1\}$.
It allows the expression of a continuous function $f$ of $m$ variables as a $2^{mL}$ dimensional tensor.
In this paper, we set $L=8$ for all three variables $m=3$ to represent Eq.~\eqref{eq:multivariable function}.
The function's heatmap concerning $(x_1, x_2) \in [0, 1)^{\otimes2}$ with $x_3 = 0.5$ is depicted in Fig.~\ref{fig:factorizing tensor}~(a).
To construct the tensor $\Psi$ of Eq.~\eqref{eq:multivariable function}, we employed a one-dimensional variable ordering, in the same order as the MPN structure shown in Fig.~\ref{fig:factorizing tensor}(b) that was generated via the sequential SVD from $\Psi$.
The authors of Ref.~\cite{tindallCompressingMultivariateFunctions2024} have shown that TTN structures designed to separate each of the three variables $(x_1, x_2, x_3)$ are reasonable, as Eq.~\eqref{eq:multivariable function} exhibits only weak correlations among them.

In our demonstration, we constructed an MPN representation of the normalized function in Eq.~\eqref{eq:multivariable function} using sequential SVD with $\chi_{\rm init} = 4$, as depicted in Fig.~\ref{fig:factorizing tensor}(b).
The resulting MPN was then passed into the TTN update methods of maximizing the fidelity with $[\chi_1, \chi_2, \chi_3] = [4, 8, 16]$.
The structural optimization was applied to MPN when $\chi_1 = 4$.
The thresholds for the convergence of fidelity and EE were set to $\epsilon_{F} = 1 \times 10^{-10}$ and $\epsilon_{\mathcal{S}} = 1 \times 10^{-14}$, respectively.

The TTN structure obtained by TTNOpt is illustrated in Fig.~\ref{fig:factorizing tensor}(c).
TTNOpt successfully identified a TTN structure that reflects the insight that the three variables are not correlated in a bit-wise way.
Table~\ref{Table_II} shows the EEs and fidelities obtained for both the MPN and the optimized TTN.
The EEs for the TTN are lower than those from the MPN at each bond dimension.
For instance, at $\chi = 16$, the EE of the MPN reaches $0.9745$, while that of the TTN remains at $0.6169$.
It should also be noted that, when comparing the fidelities of the resulting TTN and the MPN, the memory footprint of the former is bigger than that of the latter.
It should also be noted that, concerning the fidelities of the two TNs, the memory footprint of the resulting TTN exceeds that of the MPN.
For example, at $\chi=16$, the TTN requires approximately $1.65$ times more memory.
These results highlight that TTNOpt not only improves compression fidelity but also reveals efficient data sparsity based on the entanglement structure, offering a significant advantage over conventional MPN-based approaches.
\begin{table}[htbp]
\caption{The average bond EEs $\mathcal{S}$ and fidelities $ F = \left| \Psi \prod_{v\in \bm v}\circ~v^* \right|$ obtained for the optimized TTN and the fixed MPN, respectively.
The average EEs are taken over auxiliary and physical bonds and rounded to four decimal places.
On the other hand, the average fidelity is taken over auxiliary bonds and truncated after the fifth significant figure.
}
\label{Table_II}
\begin{center}
\begin{tabular}{l c c}
\hline
\hline
Type & TTN & MPN \\
\hline
$\mathcal{S}$ ~($\chi = 4$) & 0.3825 & 0.5631 \\
$\mathcal{S}$ ~($\chi = 8$)  & 0.5378 & 0.7920 \\
$\mathcal{S}$ ~($\chi = 16$)  & 0.6169 & 0.9745 \\
\hline
$F$ ($\chi = 4$) & 0.39408 & 0.33479 \\
$F$ ($\chi = 8$) & 0.72512 & 0.57523 \\
$F$ ($\chi = 16$) & 0.99997 & 0.83884 \\
\hline
\hline
\end{tabular}
\end{center}
\end{table}

\subsection{Multivariate normal distribution}\label{sec:benchmark reconstruction}

\begin{figure}[h]
  \centering
  \includegraphics[clip, width=3.3in]{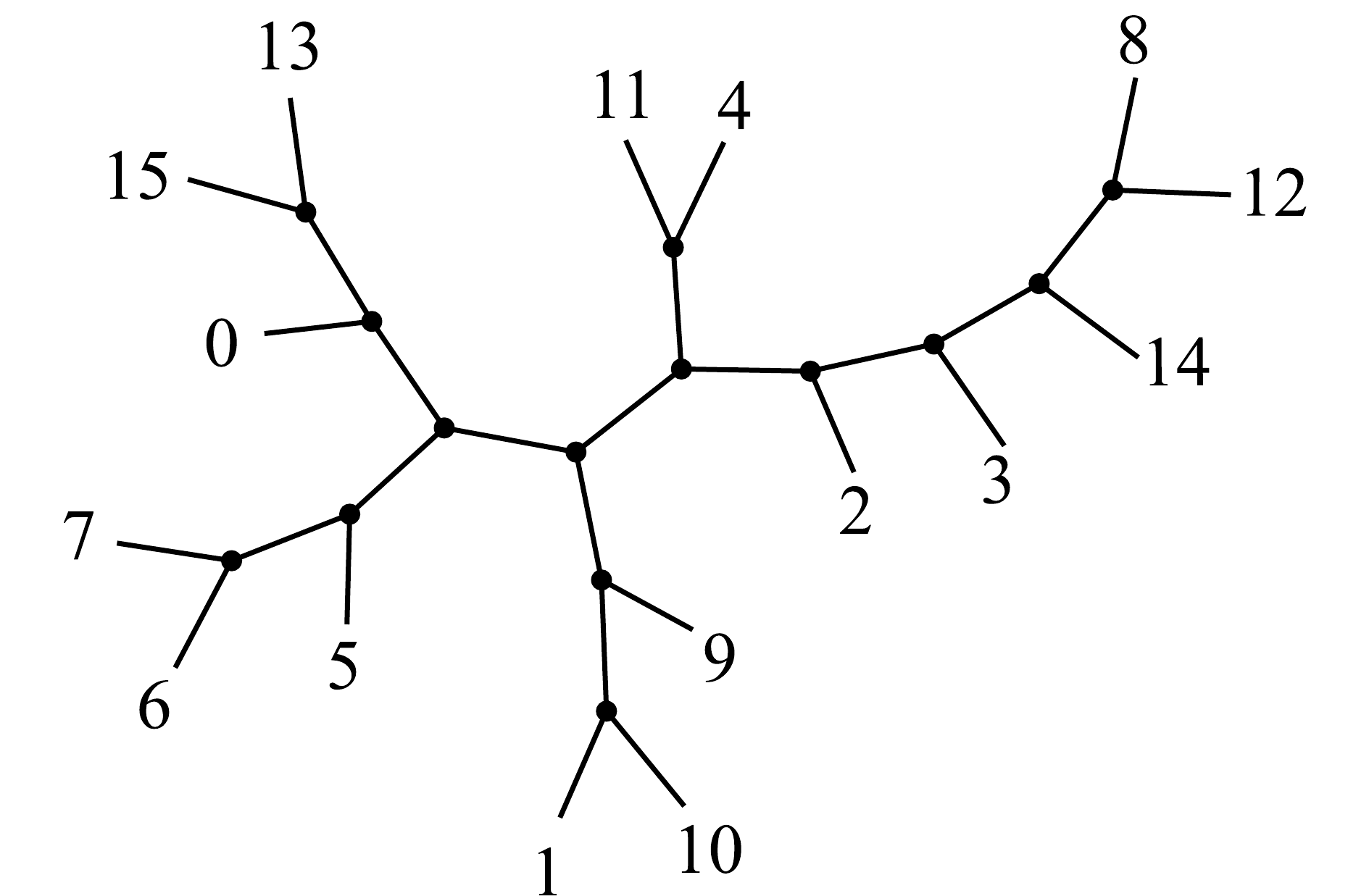}
  \caption{The tree structure to define the covariance matrix $K$ described in Fig.~\ref{fig:cov} in the numerical experiment.
  }
  \label{fig:cov_tree}
\end{figure}

Finally, as an illustrative example of TTN reconstruction, we consider the TTN representation of a multivariate normal distribution~\cite{manabeStatePreparationMultivariate2024}. 
The probability density function of a multivariate normal distribution with mean zero and covariance matrix $K$ is given as follows:
\begin{equation}\label{ eq:multivariable cov}
    f(\bm{x})=\exp\left(-\frac{1}{2}\bm{x}^{T}K^{-1}\bm{x}\right)~,
\end{equation}
where $\bm{x}$ is a $D$-dimensional vector and the normalization term is omitted for simplicity.
In our demonstration, we set $D=16$ and each component of $\bm{x}$ is discretized over the range $[-5, 5]$ using $L=4$ bits of precision.
We use the covariance matrix $K$ shown in Fig.~\ref{fig:cov}, which is constructed based on the tree structure illustrated in Fig.~\ref{fig:cov_tree}.
The elements of $K$ are defined using a parameter $\rho=0.2$ and decay exponentially with the shortest path length between the variables on the tree.
In this setting, $f(\bm{x})$ can be efficiently represented using a TTN whose geometry corresponds to the tree graph in Fig.~\ref{fig:cov_tree}.
\begin{figure*}[h]
  \centering
  \includegraphics[clip, width=0.95\textwidth]{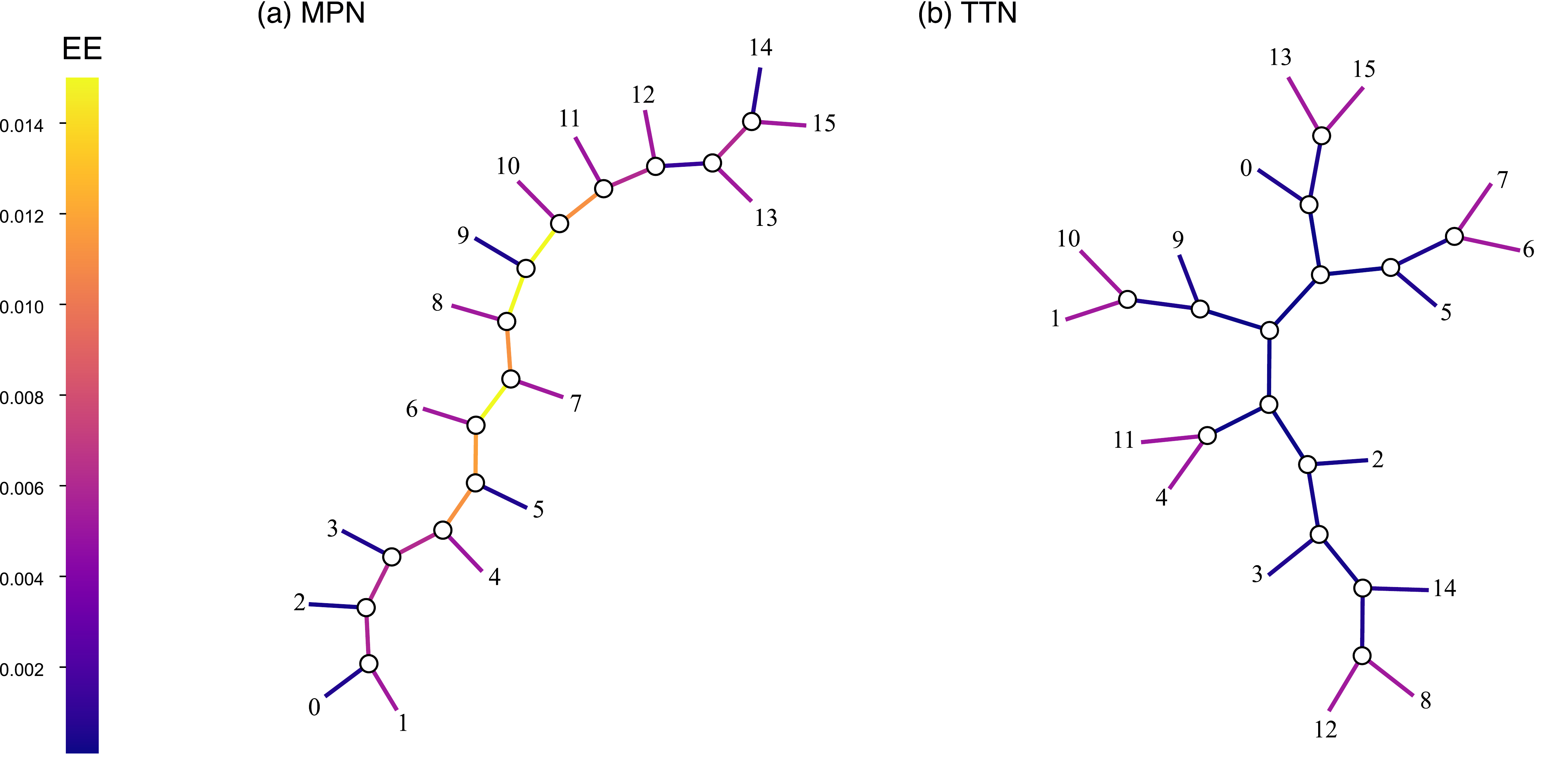}
  \caption{The initial and final structure of the tree tensor network in the reconstruction.
    The colors in the edges represent the amount of entanglement entropy.}
  \label{fig:reconstruction}
\end{figure*}

Building upon this setup, we explore whether automatic structural optimization can recover the optimal network structure of the multivariate normal distribution with tree-like correlation, following the same task as in \cite{manabeStatePreparationMultivariate2024}.
Each bare bond carries $2^4$ degrees of freedom representing $L$-bit precision.
Initially, $f(\bm{x})$ is constructed as an MPN with auxiliary bond dimension $\chi=16$ by using the Tensor Cross Interpolation (TCI) method~\cite{PhysRevX.13.021015,fernandezLearningTensorNetworks2024a,note2}.
Subsequently, the reconstruction algorithm is applied, optimizing the TTN structure based on bipartite EEs.
During the sweeps, we use the default values in $\epsilon_{\rm EE}$ and $\delta_{\rm EE}$.

The resulting TTN is shown in Fig.~\ref{fig:reconstruction}.  
The optimized structure perfectly matches the tree structure in Fig.~\ref{fig:cov_tree}, successfully capturing the underlying correlations.
Moreover, the bipartite EEs on the edges are reduced compared to those in the initial MPN.
These results demonstrate that the TTNOpt can detect hidden correlation structures of $f(\bm{x})$ and replace the MPN with the more efficient TTN representation.

\section{Summary}\label{sec:summary}
We have developed a TTN manipulation package for analyzing the ground states of quantum spin systems and general tensor data.
The TTNOpt package conducts the local structural reconnection during the sweep procedures based on two-tensor updates.
This enables us to search for effective and efficient TTN representations, surpassing the simple MPN structure.

As a demonstration, we first applied the ground state search, including the structural optimization, to the hierarchical chain model~\cite{hikiharaAutomaticStructuralOptimization2023}.
We confirmed the resulting TTN and those of EEs are consistent with those in Ref.~\cite{hikiharaAutomaticStructuralOptimization2023}.
Second, we applied the fidelity-based update method to quantic MPN, representing the three-variable function in Ref.~\cite{tindallCompressingMultivariateFunctions2024}, in both cases, with and without structural optimization.
We corroborated that a well-structured TTN can achieve better convergence in terms of fidelity when approximating the target data.
Finally, we applied TTNOpt for MPN, representing the multi-variable probability density function where the covariance matrix is explicitly defined by the tree structure~\cite{manabeStatePreparationMultivariate2024}.
As a result, the optimized TTN structure could reproduce the same structure in the covariance relation tree, and we observed a decrease in EEs compared to the MPN.

The prospects of TTNOpt are extending its scope to fermionic systems and then applying it to quantum chemistry problems.
Although molecular systems are not one-dimensional, they have been analyzed using MPN-based DMRG in many cases.
On the other hand, the potential benefits of introducing TTN, depending on molecular structures, especially in dendritic ones, have also been discussed~\cite{nakataniEfficientTreeTensor2013,murgTreeTensorNetwork2015}.
Combining our method for time evolution algorithm such as the time-dependent variational principle (TDVP) ~\cite{PhysRevLett.107.070601,PhysRevB.94.165116} could also allow for more extended time evolutions by maintaining structures with low entanglement~\cite{10.21468/SciPostPhys.8.2.024}.
It would be implemented by alternatively applying the structural search sweeps and short-time evolution.

The factorizing tensor method is used not only to reveal the entanglement structure of the target data but also, of course, to compress the data into the TTNs.
In particular, if we obtain TTN states that approximate the target quantum states, they can be converted into quantum circuits.
In this scenario, a TTN structure with smaller bond dimensions would directly reduce the circuit depth~\cite{sugawaraEmbeddingTreeTensor2025}.
It would be suitable for enhancing the usability of intermediate-sized quantum circuits~\cite{PRXQuantum.6.010320}.

The TCI algorithm~\cite{PhysRevX.13.021015,fernandezLearningTensorNetworks2024a} constructs the MPN that approximates the function $f(x)$ by accessing the sufficiently large number of input-output pairs $(\bm{x}, f(\bm{x}))$, rather than explicitly constructing the high-rank tensor $\Psi$ representing $f(\bm{x})$.
In the TCI, MPN has been used so far, while the use of TTN opens up the possibility of more efficient data representation.
The extension of TCI to TTN is feasible~\cite{tindallCompressingMultivariateFunctions2024}.
However, the integration of TTN structural optimization into the TCI framework remains a subject for future research.

\section*{Acknowledgments}
This work is partially supported by,
KAKENHI Grant Numbers JP24K06881, JP22H01171, JP21H04446, JP21H05182, JP21H05191, and JP25KJ1773 from JSPS of Japan.
We also acknowledge support from MEXT Q-LEAP Grant No.
JPMXS0120319794, and from JST COI-NEXT No. JPMJPF2014, ASPIRE No. JPMJAP2319, and CREST No.JPMJCR24I1 and JPMJCR24I3.
We thank K. Okunishi for insightful discussions and comments.
R.W. would like to thank  D. Sels, H. Shinaoka, and the Tensor Network team at the Flatiron Institute, especially J. Tindall and M. Stoudenmire, for their useful discussions.
R.W. was supported by the $\Sigma$ Doctoral Futures Research Grant Program from Osaka University.
H.M was supported by JSTCOI-NEXT program Grant Numbers JPMJPF2014.
H.U. was supported by the COE research grant in computational science from Hyogo
Prefecture and Kobe City through Foundation for Computational Science.

\bibliographystyle{apsrev4-1}
\bibliography{ref}

\begin{thebibliography}{54}%
\makeatletter
\providecommand \@ifxundefined [1]{%
 \@ifx{#1\undefined}
}%
\providecommand \@ifnum [1]{%
 \ifnum #1\expandafter \@firstoftwo
 \else \expandafter \@secondoftwo
 \fi
}%
\providecommand \@ifx [1]{%
 \ifx #1\expandafter \@firstoftwo
 \else \expandafter \@secondoftwo
 \fi
}%
\providecommand \natexlab [1]{#1}%
\providecommand \enquote  [1]{``#1''}%
\providecommand \bibnamefont  [1]{#1}%
\providecommand \bibfnamefont [1]{#1}%
\providecommand \citenamefont [1]{#1}%
\providecommand \href@noop [0]{\@secondoftwo}%
\providecommand \href [0]{\begingroup \@sanitize@url \@href}%
\providecommand \@href[1]{\@@startlink{#1}\@@href}%
\providecommand \@@href[1]{\endgroup#1\@@endlink}%
\providecommand \@sanitize@url [0]{\catcode `\\12\catcode `\$12\catcode `\&12\catcode `\#12\catcode `\^12\catcode `\_12\catcode `\%12\relax}%
\providecommand \@@startlink[1]{}%
\providecommand \@@endlink[0]{}%
\providecommand \url  [0]{\begingroup\@sanitize@url \@url }%
\providecommand \@url [1]{\endgroup\@href {#1}{\urlprefix }}%
\providecommand \urlprefix  [0]{URL }%
\providecommand \Eprint [0]{\href }%
\providecommand \doibase [0]{http://dx.doi.org/}%
\providecommand \selectlanguage [0]{\@gobble}%
\providecommand \bibinfo  [0]{\@secondoftwo}%
\providecommand \bibfield  [0]{\@secondoftwo}%
\providecommand \translation [1]{[#1]}%
\providecommand \BibitemOpen [0]{}%
\providecommand \bibitemStop [0]{}%
\providecommand \bibitemNoStop [0]{.\EOS\space}%
\providecommand \EOS [0]{\spacefactor3000\relax}%
\providecommand \BibitemShut  [1]{\csname bibitem#1\endcsname}%
\let\auto@bib@innerbib\@empty
\bibitem [{\citenamefont {TTNOpt}(2025)}]{TTNOpt}%
  \BibitemOpen
  \bibfield  {author} {\bibinfo {author} {\bibnamefont {TTNOpt}},\ }\href@noop {} {}\bibinfo {howpublished} {\url{https://github.com/Ryo-wtnb11/TTNOpt}} (\bibinfo {year} {2025})\BibitemShut {NoStop}%
\bibitem [{\citenamefont {Roberts}\ \emph {et~al.}(2019)\citenamefont {Roberts}, \citenamefont {Milsted}, \citenamefont {Ganahl}, \citenamefont {Zalcman}, \citenamefont {Fontaine}, \citenamefont {Zou}, \citenamefont {Hidary}, \citenamefont {Vidal},\ and\ \citenamefont {Leichenauer}}]{roberts2019tensornetwork}%
  \BibitemOpen
  \bibfield  {author} {\bibinfo {author} {\bibfnamefont {C.}~\bibnamefont {Roberts}}, \bibinfo {author} {\bibfnamefont {A.}~\bibnamefont {Milsted}}, \bibinfo {author} {\bibfnamefont {M.}~\bibnamefont {Ganahl}}, \bibinfo {author} {\bibfnamefont {A.}~\bibnamefont {Zalcman}}, \bibinfo {author} {\bibfnamefont {B.}~\bibnamefont {Fontaine}}, \bibinfo {author} {\bibfnamefont {Y.}~\bibnamefont {Zou}}, \bibinfo {author} {\bibfnamefont {J.}~\bibnamefont {Hidary}}, \bibinfo {author} {\bibfnamefont {G.}~\bibnamefont {Vidal}}, \ and\ \bibinfo {author} {\bibfnamefont {S.}~\bibnamefont {Leichenauer}},\ }\href@noop {} {} (\bibinfo {year} {2019}),\ \Eprint {http://arxiv.org/abs/1905.01330} {arXiv:1905.01330 [physics.comp-ph]} \BibitemShut {NoStop}%
\bibitem [{\citenamefont {Or{\'u}s}(2014)}]{orusPracticalIntroductionTensor2014}%
  \BibitemOpen
  \bibfield  {author} {\bibinfo {author} {\bibfnamefont {R.}~\bibnamefont {Or{\'u}s}},\ }\href {\doibase 10.1016/j.aop.2014.06.013} {\bibfield  {journal} {\bibinfo  {journal} {Ann. Phys.}\ }\textbf {\bibinfo {volume} {349}},\ \bibinfo {pages} {117} (\bibinfo {year} {2014})}\BibitemShut {NoStop}%
\bibitem [{\citenamefont {Okunishi}\ \emph {et~al.}(2022)\citenamefont {Okunishi}, \citenamefont {Nishino},\ and\ \citenamefont {Ueda}}]{okunishiDevelopmentsTensorNetwork2022}%
  \BibitemOpen
  \bibfield  {author} {\bibinfo {author} {\bibfnamefont {K.}~\bibnamefont {Okunishi}}, \bibinfo {author} {\bibfnamefont {T.}~\bibnamefont {Nishino}}, \ and\ \bibinfo {author} {\bibfnamefont {H.}~\bibnamefont {Ueda}},\ }\href {\doibase 10.7566/JPSJ.91.062001} {\bibfield  {journal} {\bibinfo  {journal} {J. Phys. Soc. Jpn.}\ }\textbf {\bibinfo {volume} {91}},\ \bibinfo {pages} {062001} (\bibinfo {year} {2022})}\BibitemShut {NoStop}%
\bibitem [{\citenamefont {Or{\'u}s}(2019)}]{orusTensorNetworksComplex2019a}%
  \BibitemOpen
  \bibfield  {author} {\bibinfo {author} {\bibfnamefont {R.}~\bibnamefont {Or{\'u}s}},\ }\href {\doibase 10.1038/s42254-019-0086-7} {\bibfield  {journal} {\bibinfo  {journal} {Nat. Rev. Phys.}\ }\textbf {\bibinfo {volume} {1}},\ \bibinfo {pages} {538} (\bibinfo {year} {2019})}\BibitemShut {NoStop}%
\bibitem [{\citenamefont {Lu}\ \emph {et~al.}(2024)\citenamefont {Lu}, \citenamefont {{Kan{\'a}sz-Nagy}}, \citenamefont {Kukuljan},\ and\ \citenamefont {Cirac}}]{luTensorNetworksEfficient2024}%
  \BibitemOpen
  \bibfield  {author} {\bibinfo {author} {\bibfnamefont {S.}~\bibnamefont {Lu}}, \bibinfo {author} {\bibfnamefont {M.}~\bibnamefont {{Kan{\'a}sz-Nagy}}}, \bibinfo {author} {\bibfnamefont {I.}~\bibnamefont {Kukuljan}}, \ and\ \bibinfo {author} {\bibfnamefont {J.~I.}\ \bibnamefont {Cirac}},\ }\href {\doibase 10.48550/arXiv.2103.06872} {} (\bibinfo {year} {2024}),\ \Eprint {http://arxiv.org/abs/2103.06872} {arXiv:2103.06872 [quant-ph]} \BibitemShut {NoStop}%
\bibitem [{\citenamefont {Jobst}\ \emph {et~al.}(2024)\citenamefont {Jobst}, \citenamefont {Shen}, \citenamefont {Riofr{\'i}o}, \citenamefont {Shishenina},\ and\ \citenamefont {Pollmann}}]{jobstEfficientMPSRepresentations2024}%
  \BibitemOpen
  \bibfield  {author} {\bibinfo {author} {\bibfnamefont {B.}~\bibnamefont {Jobst}}, \bibinfo {author} {\bibfnamefont {K.}~\bibnamefont {Shen}}, \bibinfo {author} {\bibfnamefont {C.~A.}\ \bibnamefont {Riofr{\'i}o}}, \bibinfo {author} {\bibfnamefont {E.}~\bibnamefont {Shishenina}}, \ and\ \bibinfo {author} {\bibfnamefont {F.}~\bibnamefont {Pollmann}},\ }\href {\doibase 10.22331/q-2024-12-03-1544} {\bibfield  {journal} {\bibinfo  {journal} {Quantum}\ }\textbf {\bibinfo {volume} {8}},\ \bibinfo {pages} {1544} (\bibinfo {year} {2024})}\BibitemShut {NoStop}%
\bibitem [{\citenamefont {Stoudenmire}\ and\ \citenamefont {Schwab}(2017)}]{stoudenmireSupervisedLearningQuantumInspired2017}%
  \BibitemOpen
  \bibfield  {author} {\bibinfo {author} {\bibfnamefont {E.~M.}\ \bibnamefont {Stoudenmire}}\ and\ \bibinfo {author} {\bibfnamefont {D.~J.}\ \bibnamefont {Schwab}},\ }\href {\doibase 10.48550/arXiv.1605.05775} {} (\bibinfo {year} {2017}),\ \Eprint {http://arxiv.org/abs/1605.05775} {arXiv:1605.05775 [stat]} \BibitemShut {NoStop}%
\bibitem [{\citenamefont {Han}\ \emph {et~al.}(2018)\citenamefont {Han}, \citenamefont {Wang}, \citenamefont {Fan}, \citenamefont {Wang},\ and\ \citenamefont {Zhang}}]{PhysRevX.8.031012}%
  \BibitemOpen
  \bibfield  {author} {\bibinfo {author} {\bibfnamefont {Z.-Y.}\ \bibnamefont {Han}}, \bibinfo {author} {\bibfnamefont {J.}~\bibnamefont {Wang}}, \bibinfo {author} {\bibfnamefont {H.}~\bibnamefont {Fan}}, \bibinfo {author} {\bibfnamefont {L.}~\bibnamefont {Wang}}, \ and\ \bibinfo {author} {\bibfnamefont {P.}~\bibnamefont {Zhang}},\ }\href {\doibase 10.1103/PhysRevX.8.031012} {\bibfield  {journal} {\bibinfo  {journal} {Phys. Rev. X}\ }\textbf {\bibinfo {volume} {8}},\ \bibinfo {pages} {031012} (\bibinfo {year} {2018})}\BibitemShut {NoStop}%
\bibitem [{\citenamefont {Cheng}\ \emph {et~al.}(2019)\citenamefont {Cheng}, \citenamefont {Wang}, \citenamefont {Xiang},\ and\ \citenamefont {Zhang}}]{PhysRevB.99.155131}%
  \BibitemOpen
  \bibfield  {author} {\bibinfo {author} {\bibfnamefont {S.}~\bibnamefont {Cheng}}, \bibinfo {author} {\bibfnamefont {L.}~\bibnamefont {Wang}}, \bibinfo {author} {\bibfnamefont {T.}~\bibnamefont {Xiang}}, \ and\ \bibinfo {author} {\bibfnamefont {P.}~\bibnamefont {Zhang}},\ }\href {\doibase 10.1103/PhysRevB.99.155131} {\bibfield  {journal} {\bibinfo  {journal} {Phys. Rev. B}\ }\textbf {\bibinfo {volume} {99}},\ \bibinfo {pages} {155131} (\bibinfo {year} {2019})}\BibitemShut {NoStop}%
\bibitem [{\citenamefont {Oseledets}(2009)}]{oseledetsApproximationMatricesLogarithmic2009}%
  \BibitemOpen
  \bibfield  {author} {\bibinfo {author} {\bibfnamefont {I.~V.}\ \bibnamefont {Oseledets}},\ }\href {\doibase 10.1134/S1064562409050056} {\bibfield  {journal} {\bibinfo  {journal} {Dokl. Math.}\ }\textbf {\bibinfo {volume} {80}},\ \bibinfo {pages} {653} (\bibinfo {year} {2009})}\BibitemShut {NoStop}%
\bibitem [{\citenamefont {Khoromskij}(2011)}]{khoromskijOdlogNQuanticsApproximation2011}%
  \BibitemOpen
  \bibfield  {author} {\bibinfo {author} {\bibfnamefont {B.~N.}\ \bibnamefont {Khoromskij}},\ }\href {\doibase 10.1007/s00365-011-9131-1} {\bibfield  {journal} {\bibinfo  {journal} {Constr. Approx.}\ }\textbf {\bibinfo {volume} {34}},\ \bibinfo {pages} {257} (\bibinfo {year} {2011})}\BibitemShut {NoStop}%
\bibitem [{\citenamefont {Nishino}\ \emph {et~al.}(2001)\citenamefont {Nishino}, \citenamefont {Hieida}, \citenamefont {Okunishi}, \citenamefont {Maeshima}, \citenamefont {Akutsu},\ and\ \citenamefont {Gendiar}}]{10.1143/PTP.105.409}%
  \BibitemOpen
  \bibfield  {author} {\bibinfo {author} {\bibfnamefont {T.}~\bibnamefont {Nishino}}, \bibinfo {author} {\bibfnamefont {Y.}~\bibnamefont {Hieida}}, \bibinfo {author} {\bibfnamefont {K.}~\bibnamefont {Okunishi}}, \bibinfo {author} {\bibfnamefont {N.}~\bibnamefont {Maeshima}}, \bibinfo {author} {\bibfnamefont {Y.}~\bibnamefont {Akutsu}}, \ and\ \bibinfo {author} {\bibfnamefont {A.}~\bibnamefont {Gendiar}},\ }\href {\doibase 10.1143/PTP.105.409} {\bibfield  {journal} {\bibinfo  {journal} {Prog. Theor. Phys.}\ }\textbf {\bibinfo {volume} {105}},\ \bibinfo {pages} {409} (\bibinfo {year} {2001})}\BibitemShut {NoStop}%
\bibitem [{\citenamefont {Vidal}(2007)}]{PhysRevLett.99.220405}%
  \BibitemOpen
  \bibfield  {author} {\bibinfo {author} {\bibfnamefont {G.}~\bibnamefont {Vidal}},\ }\href {\doibase 10.1103/PhysRevLett.99.220405} {\bibfield  {journal} {\bibinfo  {journal} {Phys. Rev. Lett.}\ }\textbf {\bibinfo {volume} {99}},\ \bibinfo {pages} {220405} (\bibinfo {year} {2007})}\BibitemShut {NoStop}%
\bibitem [{\citenamefont {Jordan}\ \emph {et~al.}(2008)\citenamefont {Jordan}, \citenamefont {Or\'us}, \citenamefont {Vidal}, \citenamefont {Verstraete},\ and\ \citenamefont {Cirac}}]{PhysRevLett.101.250602}%
  \BibitemOpen
  \bibfield  {author} {\bibinfo {author} {\bibfnamefont {J.}~\bibnamefont {Jordan}}, \bibinfo {author} {\bibfnamefont {R.}~\bibnamefont {Or\'us}}, \bibinfo {author} {\bibfnamefont {G.}~\bibnamefont {Vidal}}, \bibinfo {author} {\bibfnamefont {F.}~\bibnamefont {Verstraete}}, \ and\ \bibinfo {author} {\bibfnamefont {J.~I.}\ \bibnamefont {Cirac}},\ }\href {\doibase 10.1103/PhysRevLett.101.250602} {\bibfield  {journal} {\bibinfo  {journal} {Phys. Rev. Lett.}\ }\textbf {\bibinfo {volume} {101}},\ \bibinfo {pages} {250602} (\bibinfo {year} {2008})}\BibitemShut {NoStop}%
\bibitem [{\citenamefont {Cirac}\ \emph {et~al.}(2021)\citenamefont {Cirac}, \citenamefont {P\'erez-Garc\'{\i}a}, \citenamefont {Schuch},\ and\ \citenamefont {Verstraete}}]{RevModPhys.93.045003}%
  \BibitemOpen
  \bibfield  {author} {\bibinfo {author} {\bibfnamefont {J.~I.}\ \bibnamefont {Cirac}}, \bibinfo {author} {\bibfnamefont {D.}~\bibnamefont {P\'erez-Garc\'{\i}a}}, \bibinfo {author} {\bibfnamefont {N.}~\bibnamefont {Schuch}}, \ and\ \bibinfo {author} {\bibfnamefont {F.}~\bibnamefont {Verstraete}},\ }\href {\doibase 10.1103/RevModPhys.93.045003} {\bibfield  {journal} {\bibinfo  {journal} {Rev. Mod. Phys.}\ }\textbf {\bibinfo {volume} {93}},\ \bibinfo {pages} {045003} (\bibinfo {year} {2021})}\BibitemShut {NoStop}%
\bibitem [{\citenamefont {Vidal}(2003)}]{PhysRevLett.91.147902}%
  \BibitemOpen
  \bibfield  {author} {\bibinfo {author} {\bibfnamefont {G.}~\bibnamefont {Vidal}},\ }\href {\doibase 10.1103/PhysRevLett.91.147902} {\bibfield  {journal} {\bibinfo  {journal} {Phys. Rev. Lett.}\ }\textbf {\bibinfo {volume} {91}},\ \bibinfo {pages} {147902} (\bibinfo {year} {2003})}\BibitemShut {NoStop}%
\bibitem [{\citenamefont {Vidal}(2004)}]{PhysRevLett.93.040502}%
  \BibitemOpen
  \bibfield  {author} {\bibinfo {author} {\bibfnamefont {G.}~\bibnamefont {Vidal}},\ }\href {\doibase 10.1103/PhysRevLett.93.040502} {\bibfield  {journal} {\bibinfo  {journal} {Phys. Rev. Lett.}\ }\textbf {\bibinfo {volume} {93}},\ \bibinfo {pages} {040502} (\bibinfo {year} {2004})}\BibitemShut {NoStop}%
\bibitem [{\citenamefont {Murg}\ \emph {et~al.}(2010)\citenamefont {Murg}, \citenamefont {Verstraete}, \citenamefont {Legeza},\ and\ \citenamefont {Noack}}]{PhysRevB.82.205105}%
  \BibitemOpen
  \bibfield  {author} {\bibinfo {author} {\bibfnamefont {V.}~\bibnamefont {Murg}}, \bibinfo {author} {\bibfnamefont {F.}~\bibnamefont {Verstraete}}, \bibinfo {author} {\bibfnamefont {O.}~\bibnamefont {Legeza}}, \ and\ \bibinfo {author} {\bibfnamefont {R.~M.}\ \bibnamefont {Noack}},\ }\href {\doibase 10.1103/PhysRevB.82.205105} {\bibfield  {journal} {\bibinfo  {journal} {Phys. Rev. B}\ }\textbf {\bibinfo {volume} {82}},\ \bibinfo {pages} {205105} (\bibinfo {year} {2010})}\BibitemShut {NoStop}%
\bibitem [{\citenamefont {Larsson}(2019)}]{larssonComputingVibrationalEigenstates2019}%
  \BibitemOpen
  \bibfield  {author} {\bibinfo {author} {\bibfnamefont {H.~R.}\ \bibnamefont {Larsson}},\ }\href {\doibase 10.1063/1.5130390} {\bibfield  {journal} {\bibinfo  {journal} {J. Chem. Phys.}\ }\textbf {\bibinfo {volume} {151}},\ \bibinfo {pages} {204102} (\bibinfo {year} {2019})}\BibitemShut {NoStop}%
\bibitem [{\citenamefont {Evenbly}\ and\ \citenamefont {Vidal}(2009)}]{PhysRevB.79.144108}%
  \BibitemOpen
  \bibfield  {author} {\bibinfo {author} {\bibfnamefont {G.}~\bibnamefont {Evenbly}}\ and\ \bibinfo {author} {\bibfnamefont {G.}~\bibnamefont {Vidal}},\ }\href {\doibase 10.1103/PhysRevB.79.144108} {\bibfield  {journal} {\bibinfo  {journal} {Phys. Rev. B}\ }\textbf {\bibinfo {volume} {79}},\ \bibinfo {pages} {144108} (\bibinfo {year} {2009})}\BibitemShut {NoStop}%
\bibitem [{\citenamefont {Zaletel}\ and\ \citenamefont {Pollmann}(2020)}]{PhysRevLett.124.037201}%
  \BibitemOpen
  \bibfield  {author} {\bibinfo {author} {\bibfnamefont {M.~P.}\ \bibnamefont {Zaletel}}\ and\ \bibinfo {author} {\bibfnamefont {F.}~\bibnamefont {Pollmann}},\ }\href {\doibase 10.1103/PhysRevLett.124.037201} {\bibfield  {journal} {\bibinfo  {journal} {Phys. Rev. Lett.}\ }\textbf {\bibinfo {volume} {124}},\ \bibinfo {pages} {037201} (\bibinfo {year} {2020})}\BibitemShut {NoStop}%
\bibitem [{\citenamefont {Sauerwein}\ \emph {et~al.}(2019)\citenamefont {Sauerwein}, \citenamefont {Molnar}, \citenamefont {Cirac},\ and\ \citenamefont {Kraus}}]{PhysRevLett.123.170504}%
  \BibitemOpen
  \bibfield  {author} {\bibinfo {author} {\bibfnamefont {D.}~\bibnamefont {Sauerwein}}, \bibinfo {author} {\bibfnamefont {A.}~\bibnamefont {Molnar}}, \bibinfo {author} {\bibfnamefont {J.~I.}\ \bibnamefont {Cirac}}, \ and\ \bibinfo {author} {\bibfnamefont {B.}~\bibnamefont {Kraus}},\ }\href {\doibase 10.1103/PhysRevLett.123.170504} {\bibfield  {journal} {\bibinfo  {journal} {Phys. Rev. Lett.}\ }\textbf {\bibinfo {volume} {123}},\ \bibinfo {pages} {170504} (\bibinfo {year} {2019})}\BibitemShut {NoStop}%
\bibitem [{\citenamefont {Perez-Garcia}\ \emph {et~al.}(2007)\citenamefont {Perez-Garcia}, \citenamefont {Verstraete}, \citenamefont {Wolf},\ and\ \citenamefont {Cirac}}]{10.5555/2011832.2011833}%
  \BibitemOpen
  \bibfield  {author} {\bibinfo {author} {\bibfnamefont {D.}~\bibnamefont {Perez-Garcia}}, \bibinfo {author} {\bibfnamefont {F.}~\bibnamefont {Verstraete}}, \bibinfo {author} {\bibfnamefont {M.~M.}\ \bibnamefont {Wolf}}, \ and\ \bibinfo {author} {\bibfnamefont {J.~I.}\ \bibnamefont {Cirac}},\ }\href {https://dl.acm.org/doi/10.5555/2011832.2011833} {\bibfield  {journal} {\bibinfo  {journal} {Quantum Inf. Comput.}\ }\textbf {\bibinfo {volume} {7}},\ \bibinfo {pages} {401–430} (\bibinfo {year} {2007})}\BibitemShut {NoStop}%
\bibitem [{\citenamefont {Oseledets}(2011)}]{doi:10.1137/090752286}%
  \BibitemOpen
  \bibfield  {author} {\bibinfo {author} {\bibfnamefont {I.~V.}\ \bibnamefont {Oseledets}},\ }\href {\doibase 10.1137/090752286} {\bibfield  {journal} {\bibinfo  {journal} {SIAM J. Sci. Comput.}\ }\textbf {\bibinfo {volume} {33}},\ \bibinfo {pages} {2295} (\bibinfo {year} {2011})}\BibitemShut {NoStop}%
\bibitem [{\citenamefont {White}(1992)}]{whiteDensityMatrixFormulation1992}%
  \BibitemOpen
  \bibfield  {author} {\bibinfo {author} {\bibfnamefont {S.~R.}\ \bibnamefont {White}},\ }\href {\doibase 10.1103/PhysRevLett.69.2863} {\bibfield  {journal} {\bibinfo  {journal} {Phys. Rev. Lett.}\ }\textbf {\bibinfo {volume} {69}},\ \bibinfo {pages} {2863} (\bibinfo {year} {1992})}\BibitemShut {NoStop}%
\bibitem [{\citenamefont {White}(1993)}]{whiteDensitymatrixAlgorithmsQuantum1993}%
  \BibitemOpen
  \bibfield  {author} {\bibinfo {author} {\bibfnamefont {S.~R.}\ \bibnamefont {White}},\ }\href {\doibase 10.1103/PhysRevB.48.10345} {\bibfield  {journal} {\bibinfo  {journal} {Phys. Rev. B}\ }\textbf {\bibinfo {volume} {48}},\ \bibinfo {pages} {10345} (\bibinfo {year} {1993})}\BibitemShut {NoStop}%
\bibitem [{\citenamefont {Schollw{\"o}ck}(2011)}]{schollwockDensitymatrixRenormalizationGroup2011}%
  \BibitemOpen
  \bibfield  {author} {\bibinfo {author} {\bibfnamefont {U.}~\bibnamefont {Schollw{\"o}ck}},\ }\href {\doibase 10.1016/j.aop.2010.09.012} {\bibfield  {journal} {\bibinfo  {journal} {Ann. Phys.}\ }\textbf {\bibinfo {volume} {326}},\ \bibinfo {pages} {96} (\bibinfo {year} {2011})},\ \bibinfo {note} {january 2011 {{Special Issue}}}\BibitemShut {NoStop}%
\bibitem [{\citenamefont {Okunishi}\ \emph {et~al.}(2023)\citenamefont {Okunishi}, \citenamefont {Ueda},\ and\ \citenamefont {Nishino}}]{10.1093/ptep/ptad018}%
  \BibitemOpen
  \bibfield  {author} {\bibinfo {author} {\bibfnamefont {K.}~\bibnamefont {Okunishi}}, \bibinfo {author} {\bibfnamefont {H.}~\bibnamefont {Ueda}}, \ and\ \bibinfo {author} {\bibfnamefont {T.}~\bibnamefont {Nishino}},\ }\href {\doibase 10.1093/ptep/ptad018} {\bibfield  {journal} {\bibinfo  {journal} {Prog. Theor. Exp. Phys.}\ }\textbf {\bibinfo {volume} {2023}},\ \bibinfo {pages} {023A02} (\bibinfo {year} {2023})}\BibitemShut {NoStop}%
\bibitem [{\citenamefont {Chan}\ and\ \citenamefont {{Head-Gordon}}(2002)}]{chanHighlyCorrelatedCalculations2002}%
  \BibitemOpen
  \bibfield  {author} {\bibinfo {author} {\bibfnamefont {G.~K.-L.}\ \bibnamefont {Chan}}\ and\ \bibinfo {author} {\bibfnamefont {M.}~\bibnamefont {{Head-Gordon}}},\ }\href {\doibase 10.1063/1.1449459} {\bibfield  {journal} {\bibinfo  {journal} {J. Chem. Phys.}\ }\textbf {\bibinfo {volume} {116}},\ \bibinfo {pages} {4462} (\bibinfo {year} {2002})}\BibitemShut {NoStop}%
\bibitem [{\citenamefont {Legeza}\ and\ \citenamefont {S\'olyom}(2003)}]{PhysRevB.68.195116}%
  \BibitemOpen
  \bibfield  {author} {\bibinfo {author} {\bibfnamefont {O.}~\bibnamefont {Legeza}}\ and\ \bibinfo {author} {\bibfnamefont {J.}~\bibnamefont {S\'olyom}},\ }\href {\doibase 10.1103/PhysRevB.68.195116} {\bibfield  {journal} {\bibinfo  {journal} {Phys. Rev. B}\ }\textbf {\bibinfo {volume} {68}},\ \bibinfo {pages} {195116} (\bibinfo {year} {2003})}\BibitemShut {NoStop}%
\bibitem [{\citenamefont {Moritz}\ \emph {et~al.}(2005)\citenamefont {Moritz}, \citenamefont {Hess},\ and\ \citenamefont {Reiher}}]{moritzConvergenceBehaviorDensitymatrix2005}%
  \BibitemOpen
  \bibfield  {author} {\bibinfo {author} {\bibfnamefont {G.}~\bibnamefont {Moritz}}, \bibinfo {author} {\bibfnamefont {B.~A.}\ \bibnamefont {Hess}}, \ and\ \bibinfo {author} {\bibfnamefont {M.}~\bibnamefont {Reiher}},\ }\href {\doibase 10.1063/1.1824891} {\bibfield  {journal} {\bibinfo  {journal} {J. Chem. Phys.}\ }\textbf {\bibinfo {volume} {122}},\ \bibinfo {pages} {024107} (\bibinfo {year} {2005})}\BibitemShut {NoStop}%
\bibitem [{\citenamefont {Legeza}\ \emph {et~al.}(2015)\citenamefont {Legeza}, \citenamefont {Veis}, \citenamefont {Poves},\ and\ \citenamefont {Dukelsky}}]{PhysRevC.92.051303}%
  \BibitemOpen
  \bibfield  {author} {\bibinfo {author} {\bibfnamefont {O.}~\bibnamefont {Legeza}}, \bibinfo {author} {\bibfnamefont {L.}~\bibnamefont {Veis}}, \bibinfo {author} {\bibfnamefont {A.}~\bibnamefont {Poves}}, \ and\ \bibinfo {author} {\bibfnamefont {J.}~\bibnamefont {Dukelsky}},\ }\href {\doibase 10.1103/PhysRevC.92.051303} {\bibfield  {journal} {\bibinfo  {journal} {Phys. Rev. C}\ }\textbf {\bibinfo {volume} {92}},\ \bibinfo {pages} {051303} (\bibinfo {year} {2015})}\BibitemShut {NoStop}%
\bibitem [{\citenamefont {Li}\ \emph {et~al.}(2022)\citenamefont {Li}, \citenamefont {Ren}, \citenamefont {Yang},\ and\ \citenamefont {Shuai}}]{liFlySwappingAlgorithm2022}%
  \BibitemOpen
  \bibfield  {author} {\bibinfo {author} {\bibfnamefont {W.}~\bibnamefont {Li}}, \bibinfo {author} {\bibfnamefont {J.}~\bibnamefont {Ren}}, \bibinfo {author} {\bibfnamefont {H.}~\bibnamefont {Yang}}, \ and\ \bibinfo {author} {\bibfnamefont {Z.}~\bibnamefont {Shuai}},\ }\href {\doibase 10.1088/1361-648X/ac640e} {\bibfield  {journal} {\bibinfo  {journal} {J. Phys. Condens. Matter}\ }\textbf {\bibinfo {volume} {34}},\ \bibinfo {pages} {254003} (\bibinfo {year} {2022})}\BibitemShut {NoStop}%
\bibitem [{\citenamefont {Hikihara}\ \emph {et~al.}(2023)\citenamefont {Hikihara}, \citenamefont {Ueda}, \citenamefont {Okunishi}, \citenamefont {Harada},\ and\ \citenamefont {Nishino}}]{hikiharaAutomaticStructuralOptimization2023}%
  \BibitemOpen
  \bibfield  {author} {\bibinfo {author} {\bibfnamefont {T.}~\bibnamefont {Hikihara}}, \bibinfo {author} {\bibfnamefont {H.}~\bibnamefont {Ueda}}, \bibinfo {author} {\bibfnamefont {K.}~\bibnamefont {Okunishi}}, \bibinfo {author} {\bibfnamefont {K.}~\bibnamefont {Harada}}, \ and\ \bibinfo {author} {\bibfnamefont {T.}~\bibnamefont {Nishino}},\ }\href {\doibase 10.1103/PhysRevResearch.5.013031} {\bibfield  {journal} {\bibinfo  {journal} {Phys. Rev. Res.}\ }\textbf {\bibinfo {volume} {5}},\ \bibinfo {pages} {013031} (\bibinfo {year} {2023})}\BibitemShut {NoStop}%
\bibitem [{\citenamefont {Hikihara}\ \emph {et~al.}(2024)\citenamefont {Hikihara}, \citenamefont {Ueda}, \citenamefont {Okunishi}, \citenamefont {Harada},\ and\ \citenamefont {Nishino}}]{hikiharaVisualizationEntanglementGeometry2024}%
  \BibitemOpen
  \bibfield  {author} {\bibinfo {author} {\bibfnamefont {T.}~\bibnamefont {Hikihara}}, \bibinfo {author} {\bibfnamefont {H.}~\bibnamefont {Ueda}}, \bibinfo {author} {\bibfnamefont {K.}~\bibnamefont {Okunishi}}, \bibinfo {author} {\bibfnamefont {K.}~\bibnamefont {Harada}}, \ and\ \bibinfo {author} {\bibfnamefont {T.}~\bibnamefont {Nishino}},\ }\href {\doibase 10.48550/arXiv.2401.16000} {} (\bibinfo {year} {2024}),\ \Eprint {http://arxiv.org/abs/2401.16000} {arXiv:2401.16000 [cond-mat]} \BibitemShut {NoStop}%
\bibitem [{\citenamefont {Hikihara}\ \emph {et~al.}(2025)\citenamefont {Hikihara}, \citenamefont {Ueda}, \citenamefont {Okunishi}, \citenamefont {Harada},\ and\ \citenamefont {Nishino}}]{hikiharaImprovingAccuracyTreetensor2025}%
  \BibitemOpen
  \bibfield  {author} {\bibinfo {author} {\bibfnamefont {T.}~\bibnamefont {Hikihara}}, \bibinfo {author} {\bibfnamefont {H.}~\bibnamefont {Ueda}}, \bibinfo {author} {\bibfnamefont {K.}~\bibnamefont {Okunishi}}, \bibinfo {author} {\bibfnamefont {K.}~\bibnamefont {Harada}}, \ and\ \bibinfo {author} {\bibfnamefont {T.}~\bibnamefont {Nishino}},\ }\href {\doibase 10.48550/arXiv.2501.15514} {} (\bibinfo {year} {2025}),\ \Eprint {http://arxiv.org/abs/2501.15514} {arXiv:2501.15514 [cond-mat]} \BibitemShut {NoStop}%
\bibitem [{\citenamefont {Wilson}(1975)}]{wilsonRenormalizationGroupCritical1975}%
  \BibitemOpen
  \bibfield  {author} {\bibinfo {author} {\bibfnamefont {K.~G.}\ \bibnamefont {Wilson}},\ }\href {\doibase 10.1103/RevModPhys.47.773} {\bibfield  {journal} {\bibinfo  {journal} {Rev. Mod. Phys.}\ }\textbf {\bibinfo {volume} {47}},\ \bibinfo {pages} {773} (\bibinfo {year} {1975})}\BibitemShut {NoStop}%
\bibitem [{\citenamefont {Hikihara}\ \emph {et~al.}(1999)\citenamefont {Hikihara}, \citenamefont {Furusaki},\ and\ \citenamefont {Sigrist}}]{hikiharaNumericalRenormalizationgroupStudy1999}%
  \BibitemOpen
  \bibfield  {author} {\bibinfo {author} {\bibfnamefont {T.}~\bibnamefont {Hikihara}}, \bibinfo {author} {\bibfnamefont {A.}~\bibnamefont {Furusaki}}, \ and\ \bibinfo {author} {\bibfnamefont {M.}~\bibnamefont {Sigrist}},\ }\href {\doibase 10.1103/PhysRevB.60.12116} {\bibfield  {journal} {\bibinfo  {journal} {Phys. Rev. B}\ }\textbf {\bibinfo {volume} {60}},\ \bibinfo {pages} {12116} (\bibinfo {year} {1999})}\BibitemShut {NoStop}%
\bibitem [{not({\natexlab{a}})}]{note3}%
  \BibitemOpen
  \bibinfo {note} {Specifically, each sweep searches for the lowest-energy state with $M'$, so that the renormalized wave function after the sweep span on the subspaces labeled by $\{\cdots, M'-1, M', M'+1, \cdots\}$ as long as $\chi_{\rm{init}} > 1$.}\BibitemShut {Stop}%
\bibitem [{not({\natexlab{b}})}]{note1}%
  \BibitemOpen
  \bibinfo {note} {There should be caution about discrepancies between the isometry indices $i$ in the main text and the output format. When TTNOpt outputs the physical properties on ``basic.csv'', indices of isometries $i \in [0, N_{\rm t}-1]$ are shifted by adding the system size $N$ to assign physical sites as nodes with index from $[0, N-1]$.}\BibitemShut {Stop}%
\bibitem [{\citenamefont {Ma}\ \emph {et~al.}(1979)\citenamefont {Ma}, \citenamefont {Dasgupta},\ and\ \citenamefont {Hu}}]{PhysRevLett.43.1434}%
  \BibitemOpen
  \bibfield  {author} {\bibinfo {author} {\bibfnamefont {S.-k.}\ \bibnamefont {Ma}}, \bibinfo {author} {\bibfnamefont {C.}~\bibnamefont {Dasgupta}}, \ and\ \bibinfo {author} {\bibfnamefont {C.-k.}\ \bibnamefont {Hu}},\ }\href {\doibase 10.1103/PhysRevLett.43.1434} {\bibfield  {journal} {\bibinfo  {journal} {Phys. Rev. Lett.}\ }\textbf {\bibinfo {volume} {43}},\ \bibinfo {pages} {1434} (\bibinfo {year} {1979})}\BibitemShut {NoStop}%
\bibitem [{\citenamefont {Tindall}\ \emph {et~al.}(2024)\citenamefont {Tindall}, \citenamefont {Stoudenmire},\ and\ \citenamefont {Levy}}]{tindallCompressingMultivariateFunctions2024}%
  \BibitemOpen
  \bibfield  {author} {\bibinfo {author} {\bibfnamefont {J.}~\bibnamefont {Tindall}}, \bibinfo {author} {\bibfnamefont {M.}~\bibnamefont {Stoudenmire}}, \ and\ \bibinfo {author} {\bibfnamefont {R.}~\bibnamefont {Levy}},\ }\href {\doibase 10.48550/arXiv.2410.03572} {} (\bibinfo {year} {2024}),\ \Eprint {http://arxiv.org/abs/2410.03572} {arXiv:2410.03572} \BibitemShut {NoStop}%
\bibitem [{\citenamefont {Manabe}\ and\ \citenamefont {Sano}(2024)}]{manabeStatePreparationMultivariate2024}%
  \BibitemOpen
  \bibfield  {author} {\bibinfo {author} {\bibfnamefont {H.}~\bibnamefont {Manabe}}\ and\ \bibinfo {author} {\bibfnamefont {Y.}~\bibnamefont {Sano}},\ }\href {\doibase 10.48550/arXiv.2412.12067} {} (\bibinfo {year} {2024}),\ \Eprint {http://arxiv.org/abs/2412.12067} {arXiv:2412.12067 [quant-ph]} \BibitemShut {NoStop}%
\bibitem [{\citenamefont {Shinaoka}\ \emph {et~al.}(2023)\citenamefont {Shinaoka}, \citenamefont {Wallerberger}, \citenamefont {Murakami}, \citenamefont {Nogaki}, \citenamefont {Sakurai}, \citenamefont {Werner},\ and\ \citenamefont {Kauch}}]{PhysRevX.13.021015}%
  \BibitemOpen
  \bibfield  {author} {\bibinfo {author} {\bibfnamefont {H.}~\bibnamefont {Shinaoka}}, \bibinfo {author} {\bibfnamefont {M.}~\bibnamefont {Wallerberger}}, \bibinfo {author} {\bibfnamefont {Y.}~\bibnamefont {Murakami}}, \bibinfo {author} {\bibfnamefont {K.}~\bibnamefont {Nogaki}}, \bibinfo {author} {\bibfnamefont {R.}~\bibnamefont {Sakurai}}, \bibinfo {author} {\bibfnamefont {P.}~\bibnamefont {Werner}}, \ and\ \bibinfo {author} {\bibfnamefont {A.}~\bibnamefont {Kauch}},\ }\href {\doibase 10.1103/PhysRevX.13.021015} {\bibfield  {journal} {\bibinfo  {journal} {Phys. Rev. X}\ }\textbf {\bibinfo {volume} {13}},\ \bibinfo {pages} {021015} (\bibinfo {year} {2023})}\BibitemShut {NoStop}%
\bibitem [{\citenamefont {Fern{\'a}ndez}\ \emph {et~al.}(2024)\citenamefont {Fern{\'a}ndez}, \citenamefont {Ritter}, \citenamefont {Jeannin}, \citenamefont {Li}, \citenamefont {Kloss}, \citenamefont {Louvet}, \citenamefont {Terasaki}, \citenamefont {Parcollet}, \citenamefont {von Delft}, \citenamefont {Shinaoka},\ and\ \citenamefont {Waintal}}]{fernandezLearningTensorNetworks2024a}%
  \BibitemOpen
  \bibfield  {author} {\bibinfo {author} {\bibfnamefont {Y.~N.}\ \bibnamefont {Fern{\'a}ndez}}, \bibinfo {author} {\bibfnamefont {M.~K.}\ \bibnamefont {Ritter}}, \bibinfo {author} {\bibfnamefont {M.}~\bibnamefont {Jeannin}}, \bibinfo {author} {\bibfnamefont {J.-W.}\ \bibnamefont {Li}}, \bibinfo {author} {\bibfnamefont {T.}~\bibnamefont {Kloss}}, \bibinfo {author} {\bibfnamefont {T.}~\bibnamefont {Louvet}}, \bibinfo {author} {\bibfnamefont {S.}~\bibnamefont {Terasaki}}, \bibinfo {author} {\bibfnamefont {O.}~\bibnamefont {Parcollet}}, \bibinfo {author} {\bibfnamefont {J.}~\bibnamefont {von Delft}}, \bibinfo {author} {\bibfnamefont {H.}~\bibnamefont {Shinaoka}}, \ and\ \bibinfo {author} {\bibfnamefont {X.}~\bibnamefont {Waintal}},\ }\href {\doibase 10.48550/arXiv.2407.02454} {} (\bibinfo {year} {2024}),\ \Eprint {http://arxiv.org/abs/2407.02454} {arXiv:2407.02454 [physics]} \BibitemShut {NoStop}%
\bibitem [{not({\natexlab{c}})}]{note2}%
  \BibitemOpen
  \bibinfo {note} {It is worth mentioning that the MPN constructed by the TCI must be transformed into the mixed canonical MPN, since it has the pivots on auxiliary bonds and does not follow the rule of TTNOpt as an input TTN. We put the TCI code in ``sample/reconstruct'' in TTNOpt such that users can use for broad data.}\BibitemShut {Stop}%
\bibitem [{\citenamefont {Nakatani}\ and\ \citenamefont {Chan}(2013)}]{nakataniEfficientTreeTensor2013}%
  \BibitemOpen
  \bibfield  {author} {\bibinfo {author} {\bibfnamefont {N.}~\bibnamefont {Nakatani}}\ and\ \bibinfo {author} {\bibfnamefont {G.~K.-L.}\ \bibnamefont {Chan}},\ }\href {\doibase 10.1063/1.4798639} {\bibfield  {journal} {\bibinfo  {journal} {J. Chem. Phys.}\ }\textbf {\bibinfo {volume} {138}},\ \bibinfo {pages} {134113} (\bibinfo {year} {2013})}\BibitemShut {NoStop}%
\bibitem [{\citenamefont {Murg}\ \emph {et~al.}(2015)\citenamefont {Murg}, \citenamefont {Verstraete}, \citenamefont {Schneider}, \citenamefont {Nagy},\ and\ \citenamefont {Legeza}}]{murgTreeTensorNetwork2015}%
  \BibitemOpen
  \bibfield  {author} {\bibinfo {author} {\bibfnamefont {V.}~\bibnamefont {Murg}}, \bibinfo {author} {\bibfnamefont {F.}~\bibnamefont {Verstraete}}, \bibinfo {author} {\bibfnamefont {R.}~\bibnamefont {Schneider}}, \bibinfo {author} {\bibfnamefont {P.~R.}\ \bibnamefont {Nagy}}, \ and\ \bibinfo {author} {\bibfnamefont {{\"O}.}~\bibnamefont {Legeza}},\ }\href {\doibase 10.1021/ct501187j} {\bibfield  {journal} {\bibinfo  {journal} {J. Chem. Theory Comput.}\ }\textbf {\bibinfo {volume} {11}},\ \bibinfo {pages} {1027} (\bibinfo {year} {2015})}\BibitemShut {NoStop}%
\bibitem [{\citenamefont {Haegeman}\ \emph {et~al.}(2011)\citenamefont {Haegeman}, \citenamefont {Cirac}, \citenamefont {Osborne}, \citenamefont {Pi\ifmmode~\check{z}\else \v{z}\fi{}orn}, \citenamefont {Verschelde},\ and\ \citenamefont {Verstraete}}]{PhysRevLett.107.070601}%
  \BibitemOpen
  \bibfield  {author} {\bibinfo {author} {\bibfnamefont {J.}~\bibnamefont {Haegeman}}, \bibinfo {author} {\bibfnamefont {J.~I.}\ \bibnamefont {Cirac}}, \bibinfo {author} {\bibfnamefont {T.~J.}\ \bibnamefont {Osborne}}, \bibinfo {author} {\bibfnamefont {I.}~\bibnamefont {Pi\ifmmode~\check{z}\else \v{z}\fi{}orn}}, \bibinfo {author} {\bibfnamefont {H.}~\bibnamefont {Verschelde}}, \ and\ \bibinfo {author} {\bibfnamefont {F.}~\bibnamefont {Verstraete}},\ }\href {\doibase 10.1103/PhysRevLett.107.070601} {\bibfield  {journal} {\bibinfo  {journal} {Phys. Rev. Lett.}\ }\textbf {\bibinfo {volume} {107}},\ \bibinfo {pages} {070601} (\bibinfo {year} {2011})}\BibitemShut {NoStop}%
\bibitem [{\citenamefont {Haegeman}\ \emph {et~al.}(2016)\citenamefont {Haegeman}, \citenamefont {Lubich}, \citenamefont {Oseledets}, \citenamefont {Vandereycken},\ and\ \citenamefont {Verstraete}}]{PhysRevB.94.165116}%
  \BibitemOpen
  \bibfield  {author} {\bibinfo {author} {\bibfnamefont {J.}~\bibnamefont {Haegeman}}, \bibinfo {author} {\bibfnamefont {C.}~\bibnamefont {Lubich}}, \bibinfo {author} {\bibfnamefont {I.}~\bibnamefont {Oseledets}}, \bibinfo {author} {\bibfnamefont {B.}~\bibnamefont {Vandereycken}}, \ and\ \bibinfo {author} {\bibfnamefont {F.}~\bibnamefont {Verstraete}},\ }\href {\doibase 10.1103/PhysRevB.94.165116} {\bibfield  {journal} {\bibinfo  {journal} {Phys. Rev. B}\ }\textbf {\bibinfo {volume} {94}},\ \bibinfo {pages} {165116} (\bibinfo {year} {2016})}\BibitemShut {NoStop}%
\bibitem [{\citenamefont {Bauernfeind}\ and\ \citenamefont {Aichhorn}(2020)}]{10.21468/SciPostPhys.8.2.024}%
  \BibitemOpen
  \bibfield  {author} {\bibinfo {author} {\bibfnamefont {D.}~\bibnamefont {Bauernfeind}}\ and\ \bibinfo {author} {\bibfnamefont {M.}~\bibnamefont {Aichhorn}},\ }\href {\doibase 10.21468/SciPostPhys.8.2.024} {\bibfield  {journal} {\bibinfo  {journal} {SciPost Phys.}\ }\textbf {\bibinfo {volume} {8}},\ \bibinfo {pages} {024} (\bibinfo {year} {2020})}\BibitemShut {NoStop}%
\bibitem [{\citenamefont {Sugawara}\ \emph {et~al.}(2025)\citenamefont {Sugawara}, \citenamefont {Inomata}, \citenamefont {Okubo},\ and\ \citenamefont {Todo}}]{sugawaraEmbeddingTreeTensor2025}%
  \BibitemOpen
  \bibfield  {author} {\bibinfo {author} {\bibfnamefont {S.}~\bibnamefont {Sugawara}}, \bibinfo {author} {\bibfnamefont {K.}~\bibnamefont {Inomata}}, \bibinfo {author} {\bibfnamefont {T.}~\bibnamefont {Okubo}}, \ and\ \bibinfo {author} {\bibfnamefont {S.}~\bibnamefont {Todo}},\ }\href {\doibase 10.48550/arXiv.2501.18856} {} (\bibinfo {year} {2025}),\ \Eprint {http://arxiv.org/abs/2501.18856} {arXiv:2501.18856 [quant-ph]} \BibitemShut {NoStop}%
\bibitem [{\citenamefont {Schuhmacher}\ \emph {et~al.}(2025)\citenamefont {Schuhmacher}, \citenamefont {Ballarin}, \citenamefont {Baiardi}, \citenamefont {Magnifico}, \citenamefont {Tacchino}, \citenamefont {Montangero},\ and\ \citenamefont {Tavernelli}}]{PRXQuantum.6.010320}%
  \BibitemOpen
  \bibfield  {author} {\bibinfo {author} {\bibfnamefont {J.}~\bibnamefont {Schuhmacher}}, \bibinfo {author} {\bibfnamefont {M.}~\bibnamefont {Ballarin}}, \bibinfo {author} {\bibfnamefont {A.}~\bibnamefont {Baiardi}}, \bibinfo {author} {\bibfnamefont {G.}~\bibnamefont {Magnifico}}, \bibinfo {author} {\bibfnamefont {F.}~\bibnamefont {Tacchino}}, \bibinfo {author} {\bibfnamefont {S.}~\bibnamefont {Montangero}}, \ and\ \bibinfo {author} {\bibfnamefont {I.}~\bibnamefont {Tavernelli}},\ }\href {\doibase 10.1103/PRXQuantum.6.010320} {\bibfield  {journal} {\bibinfo  {journal} {PRX Quantum}\ }\textbf {\bibinfo {volume} {6}},\ \bibinfo {pages} {010320} (\bibinfo {year} {2025})}\BibitemShut {NoStop}%
\end{thebibliography}%
\end{document}